\documentclass[showpacs,preprintnumbers,amsmath,amssymb,APSl,prd,nofootinbib,superscriptaddress, twocolumn]{revtex4-1}

\usepackage{dcolumn}
\usepackage{bm}
\usepackage{ifpdf}
\usepackage{hyperref}
\usepackage{bm}
\usepackage{xcolor,color,graphicx,graphics}
\usepackage[spanish,english]{babel}
\usepackage[latin1]{inputenc}
\usepackage[OT1]{fontenc}
\usepackage{latexsym,amssymb,amsmath,amsfonts}
\usepackage{makeidx}
\usepackage{epsfig,subfigure}
\usepackage{natbib}
\usepackage{epstopdf}
\usepackage{mathrsfs}
\hypersetup{colorlinks=true, linkcolor=blue, citecolor=green}
\usepackage{enumerate}
\usepackage{xcolor}
 \usepackage{multirow}

\usepackage{ulem}

\definecolor{red}{rgb}{1,0,0}

\def\+{^\dagger}

\def\<{\leftarrow}
\def\>{\rightarrow}

\def\({\left(}
\def\){\right)}

\newcommand{\bi}{\begin{itemize}} 				\newcommand{\ei}{\end{itemize}}
\newcommand{\benu}{\begin{enumerate}} 		\newcommand{\enu}{\end{enumerate}}
\newcommand{\bd}{\begin{dinglist}{0}}     \newcommand{\ed}{\end{dinglist}}
\newcommand{\bfig}{\begin{figure}[htbp]}  \newcommand{\efig}{\end{figure}}
        			
\newcommand{\bc}{\begin{center}} 				  \newcommand{\ec}{\end{center}}
\newcommand{\be}{\begin{equation}} 				\newcommand{\ee}{\end{equation}}
\newcommand{\bsub}{\begin{subequations}}  \newcommand{\esub}{\end{subequations}}
\newcommand{\ben}{\begin{eqnarray}} 			\newcommand{\een}{\end{eqnarray}}
\newcommand{\ba}[1]{\begin{array}{#1}} 		\newcommand{\ea}{\end{array}}
\newcommand{\bea}{\begin{equation}\begin{array}{rcl}}
\newcommand{\eea}{\end{array}\end{equation}}

\begin{document}

\title{Light ring images of double photon spheres in black hole and wormhole space-times}

\author{Merce Guerrero} \email{merguerr@ucm.es}
\affiliation{Departamento de F\'isica Te\'orica and IPARCOS,
	Universidad Complutense de Madrid, E-28040 Madrid, Spain}
	\author{Gonzalo J. Olmo} \email{gonzalo.olmo@uv.es}
\affiliation{Departamento de F\'{i}sica Te\'{o}rica and IFIC, Centro Mixto Universidad de Valencia - CSIC.
Universidad de Valencia, Burjassot-46100, Valencia, Spain}
\affiliation{Departamento de F\'isica, Universidade Federal da
Para\'\i ba, 58051-900 Jo\~ao Pessoa, Para\'\i ba, Brazil}
\author{Diego Rubiera-Garcia} \email{drubiera@ucm.es}
\affiliation{Departamento de F\'isica Te\'orica and IPARCOS,
	Universidad Complutense de Madrid, E-28040 Madrid, Spain}
	\author{Diego S\'aez-Chill\'on G\'omez} \email{diego.saez@uva.es}
\affiliation{Department of Theoretical Physics, Atomic and Optics, Campus Miguel Delibes, \\
University of Valladolid UVA, Paseo Bel\'en, 7, 47011 - Valladolid, Spain}

\date{\today}
\begin{abstract}
The silhouette of a black hole having a critical curve (an unstable bound photon orbit) when illuminated by an optically thin accretion disk whose emission is confined to the equatorial plane shows a distinctive central brightness depression (the shadow) whose outer edge consists of a series of strongly lensed, self-similar rings superimposed with the disk's direct emission. While the size and shape of the critical curve depend only on the background geometry, the pattern of bright and dark regions (including the size and depth of the shadow itself) in the image is strongly influenced by the (astro)physics of the accretion disk. This aspect makes it difficult to extract clean and clear observational discriminators between the Kerr black hole and other compact objects. In the presence of a second critical curve, however, observational differences become apparent. In this work we shall consider some spherically symmetric black hole and wormhole geometries characterized by the presence of a second critical curve, via a uniparametric family of extensions of the Schwarzschild metric. By assuming three toy models of geometrically thin accretion disks, we show the presence of additional light rings in the intermediate region between the two critical curves. The observation of such rings could represent a compelling evidence for the existence of black hole mimickers having multiple critical curves.

\end{abstract}

\maketitle

\section{Introduction}

The combination of the unicity theorems and the no-hair conjecture leads to the robust result that the most general axisymmetric electrovacuum solution of Einstein field equations of General Relativity (GR) is given by the Kerr-Newman family \cite{Robinson:1975bv}, characterized by mass, angular momentum, and charge. Despite the latter might have some effects \cite{Zajacek:2019kla}, it is typically neglected in astrophysical scenarios, where the Kerr hypothesis states that all black holes in the universe generated out of gravitational collapse (and perhaps of primordial cosmological perturbations as well \cite{Hawking:1971ei}) will therefore be described by such a solution. However, from a theoretical point of view, behind their event horizons an unpleasant surprise lurks to the careless traveller: the presence of a space-time singularity, marked by the presence of (at least) one incomplete geodesic \cite{Senovilla:2014gza}, where GR predictability and determinism are seriously threatened \cite{Curiel}.

The Kerr hypothesis has become a common test-bed for our ideas on the gravitational interaction \cite{Psaltis:2020ctj}. Since its verification somewhat resembles the Scottish black sheep tale - since it would require to check that {\it every} black hole in the universe is actually of Kerr type -, this half of the dilemma can only aspire to perform null-tests, i.e, whether a given observation matches the Kerr expectation, and to what degree of precision \cite{Abramowicz:2002vt,Glampedakis:2021oie,Shashank:2021giy}. This can be systematically implemented in a theory-agnostic way via parameterized deviations with respect to such a solution \cite{Johannsen:2011dh,Konoplya:2016jvv}. The second half would simply require to find {\it a single} unmistakable observation of an alternative compact object disguised as a Kerr black hole, i.e., a black hole mimicker \cite{Johnson-Mcdaniel:2018cdu}. A plethora of such objects has been proposed in the literature, including non-Kerr black holes \cite{Simpson:2021dyo}, boson stars \cite{Liebling:2012fv}, gravastars \cite{Mazur:2001fv}, fuzzballs \cite{Mathur:2005zp}, wormholes \cite{VisserBook}, and so on. In order to reveal the nature of black hole candidates, the newly-born field of multimessenger astronomy uses its full weaponry: electromagnetic radiation, gravitational waves, and neutrinos \cite{Cardoso:2019rvt,Addazi:2021xuf}.

Historically, the gravitational deflection of light around massive bodies has been a powerful tool in the verification of GR predictions \cite{SchneiderBook}. Since light (photons) travels along null geodesics of the background metric, the integration of the geodesic equation provides useful information to unveil the nature of any body compact enough to hold a {\it critical curve}. This is an unstable circular orbit for massless particles, and corresponds to the backtrack of the light ray trajectory from the observer's screen that asymptotically approaches a bound orbit \cite{Cunha:2018acu}. A light ray precisely on such a curve will turn an arbitrarily large number of times before any small perturbation will either make it escape to asymptotic infinity or hit the object (and be shallowed by the event horizon if the object is a black hole \cite{Bozza:2002zj}). In astrophysically relevant scenarios, a black hole (or a compact enough object) will be typically surrounded by its accretion disk, which provides the main source of illumination \cite{Luminet:1979nyg}.  It turns out that, while the size and shape of the critical curve is determined by the background geometry alone, the optical appearance of a compact object is heavily influenced by the astrophysics of the accretion disk. If the disk is optically thin (i.e., transparent to its own radiation), one expects the optical appearance to be dominated by a central brightness depression - {\it the shadow} -, surrounded by strongly-lensed light rays that have turned several times around the black hole near the critical curve, and which appear superimposed with the direct emission near the edge of the shadow \cite{Falcke:1999pj}.

In spherical accretion models the central depression entirely fills the inner region such that its outer edge precisely coincides with the critical curve \cite{Narayan:2019imo}. However, in geometrically thin accretion models such that the inner edge of the disk extends all the way down to the black hole horizon, there will be emission from inside the critical curve up to a certain region whose internal part has been termed as the {\it inner shadow} \cite{Chael:2021rjo}. As shown in \cite{Gralla:2019xty} the luminosity of the image in optically and geometrically thin accretion disk models will be largely dominated by the direct emission (intersecting the front side of the disk once), upon which an infinite series of self-similar rings are stacked with exponentially-suppressed contributions to the total luminosity of the object. The rough details of this picture were beautifully confirmed in 2019 by the EHT Collaboration \cite{Akiyama:2019cqa}, where the tracking of the superheated plasma in orbit around the supermassive central object of the M87 galaxy and modulated by strong magnetic fields, revealed a blurred bright ring-shaped lump of strongly bended light limiting a central black region\footnote{More precisely, this observation is consistent with the Kerr prediction for the size of the shadow and the impact parameter of the critical curve, $b_c=3\sqrt{3}M$, to a $\sim 17\%$ at $68\%$ confidence level. This fact has been argued to be able to put  modified Kerr-like metrics to a test, see e.g. \cite{Wei:2021lku}, though extracting quantitative constraints  from the direct image alone is subject to important  conceptual limitations \cite{Wielgus:2021peu}.}.

In view of the discussion above, one of the main challenges of the community is to find black hole mimickers  that, while being compatible with those tests that the Kerr solution passes (such as X-ray observations \cite{Bambi:2017iyh} and gravitational waves), it also allows for sufficiently large deviations from it to be distinguishable via observations \cite{Psaltis:2020lvx}. However, many of the observationally viable models proposed so far amount to mild modifications in the shape and size of the light rings and/or the shadow that are hard to distinguish from the Kerr one \cite{Held:2019xde,Shaikh:2019hbm,Paul:2019trt,Zeng:2020dco,Zeng:2020vsj,He:2021htq,Qin:2020xzu,Li:2021riw,Shaikh:2021cvl,Gan:2021pwu,Eichhorn:2021iwq,Guerrero:2021ues,Liu:2021yev,Okyay:2021nnh,Afrin:2021wlj,Zeng:2021mok,Guo:2021wid,Gan:2021xdl} (for a review of analytical studies see \cite{Perlick:2021aok}). Furthermore, it is extremely difficult to disentangle the degeneracy in the shadow's form between the background geometry, the modelling of the accretion flow, and the calibration parameters such as mass and angular momentum \cite{Lara:2021zth}. In this sense, a promising category of observational discriminators are those inducing qualitatively new features not found in the Kerr solution. Among them, and focusing on spherically symmetric geometries\footnote{This constraint is justified on the grounds that deviations from circularity of the shadow at near-extremal spin values are below $\lesssim 7\%$ \cite{Psaltis}, thus being weak enough to be blurred in the uncertainties of the modelling of the accretion flow.}, we find the presence of two critical curves, since they might induce additional concentric light rings around the central brightness depression \cite{Olmo:2021piq}. Such double critical curves may arise, for instance, in reflection-asymmetric wormhole solutions \cite{Wielgus:2020uqz,Wang:2020emr,Guerrero:2021pxt,Peng:2021osd}, or in models having an absolute maximum of the effective potential at the center \cite{Shaikh:2019jfr,Shaikh:2019itn}. The latter class of models admits different implementations like the one of the family of black bounce-type solutions found in \cite{Lobo:2020ffi}, whose strong deflection limit was recently discussed in \cite{Tsukamoto:2021caq}.

The main aim of this work is to study the optical appearance in terms of (new) light rings and shadows of the above family. The interest on the latter comes from different fronts: i) it interpolates between the Schwarzschild solution, a family of two-horizon black hole configurations, an extreme black hole, and a family of traversable (i.e. horizonless) wormhole configurations; ii) for a far away observer, the metric for all the possible configurations seems exactly the same;  iii) the radial function has a bounce at the center in all cases, allowing to restore the geodesically complete character of the space-time; iv) curvature is everywhere regular for non-vanishing values of the model's parameter; v) a second critical curve is present at the center of all solutions; vi) in a sub-family of the traversable wormhole configurations both critical curves are accessible for light rays travelling around any of them to reach the asymptotic observer.  From an observational perspective, the existence of two critical curves is the most succulent feature of these configurations, since they induce a new set of light rings not present in the Schwarzschild solution of GR. Hence, the analysis of the optical appearance for the different space-time configurations is the main novelty of the present work.

This work is organized as follows:  in Sec. \ref{sec:II} we introduce the family of spherically symmetric geometries and characterize the different configurations with particular emphasis on those having two accessible critical curves. A ray-tracing procedure is employed in Sec. \ref{sec:Ray} to classify the different light ray trajectories indexed by the number of (half-)orbits around the black hole and traversable wormhole configurations. This information is fed into the three canonical toy-models of geometrically thin accretion disks in Sec. \ref{Sec:accdisk} to produce the optical appearance of these two classes of configurations. Finally, in Sec. \ref{sec:V} we discuss the main highlights regarding the presence of additional light rings driven by the two critical curves and the feasibility of their detection.

\section{Generalized black bounces} \label{sec:II}

Let us consider a static, spherically symmetric geometry described by the following line element
\begin{equation}
ds^2=-A(x)dt^2+B(x)dx^2+r^2(x)d\Omega^2 \ ,
\end{equation}
where the radial coordinate $x$ spans the entire real line, $x \in (-\infty,+\infty)$, while in the black bounce-type geometry introduced in \cite{Lobo:2020ffi} the metric functions are given by
\begin{equation} \label{eq:SSS}
A(x)=B^{-1}(x)=1-\frac{2Mx^2}{(x^2+a^2)^{3/2}} \hspace{0.02cm} ; \hspace{0.02cm}
r^2(x)=x^2+a^2 \ .
\end{equation}
The behaviour of the radial function in this expression guarantees the extensibility of geodesics beyond $x=0$, thanks to the fact that at this point the area of the two-spheres $S=4\pi r^2(0)=4\pi a^2$ is finite. This bounce in the radial function can be interpreted as signal of the presence of a wormhole throat, separating the two asymptotically flat space-times, $x_{-} \in(-\infty,0),x_{+} \in(0,+\infty)$. In the limit $a \to 0$ the wormhole throat closes and the above solution reduces to the Schwarzschild one of GR,  $A(x) \approx 1-\tfrac{2M}{r}$. In the asymptotic limit, $x \to \pm \infty$, we also recover the Schwarzschild solution, while in the limit $x \to 0$ one finds instead a de Sitter core-type behaviour  $A(x) \approx 1-\tfrac{2M}{a^3}x^2$, a usual mechanism invoked in the literature to prevent the divergence of curvature scalars \cite{Ansoldi:2008jw,Maeda:2021jdc}. In this sense, it is worth pointing out that (\ref{eq:SSS}) is actually an extension of the well known Bardeen solution \cite{Bardeen} by allowing a non-trivial dependence in the radial function $r^2(x)$. Though field sources for this kind of black-bounce space-times can be sought for \cite{Bronnikov:2021uta}, for the sake of this work we shall take it as a toy model in a theory-agnostic way to illustrate the main new features brought by the presence of multiple critical curves in a shadow observation.

\begin{figure}[t!]
\includegraphics[width=8.0cm,height=5.5cm]{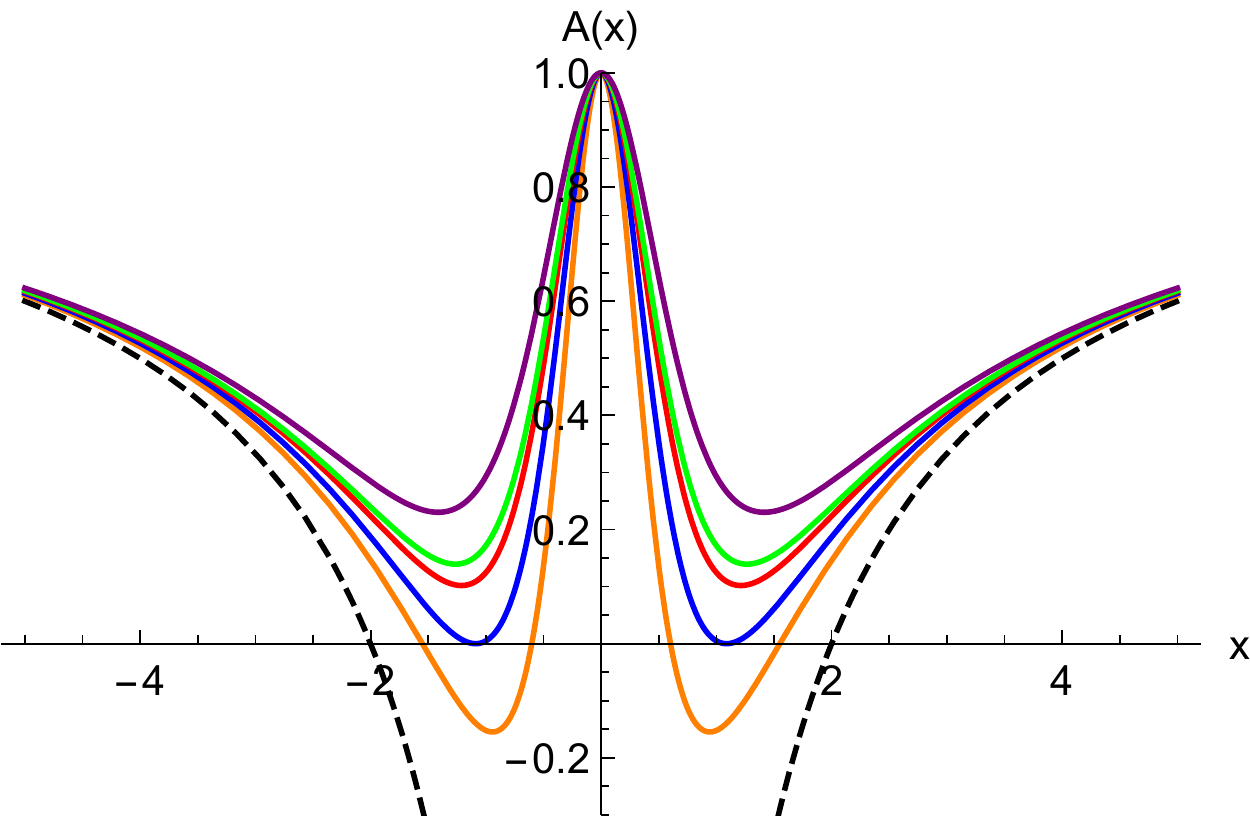}
\caption{The metric function $A(x)$ (in units of $M=1$) for the configurations with $a=0$ (Schwarzschild, dashed black), $a=2/3$ (orange), $a=\tfrac{4\sqrt{3}}{9}$ (blue), $a=6/7$ (red), $a=\tfrac{2\sqrt{5}}{5}$ (green) and $a=1$ (purple). All solutions are asymptotically flat and converge to the Schwarzschild solution of GR on that limit.}
\label{fig:metric}
\end{figure}

In order to classify the different configurations within this family let us consider first the number and location of the horizons, as given by $g^{\mu\nu}\partial_{\mu} S(x)\partial_{\nu} S(x)=0$, where $S(x)=x-x_h$, leading to $g^{xx}=A(x_h)=0$. The corresponding expression is analytical, but quite cumbersome, and for our purposes we do not need to explicitly write it here. Instead, we just need to know that two horizons arise on each side of the throat ($x=0$) as far as the condition $0<\tfrac{a}{M}<\tfrac{4\sqrt{3}}{9}$ is met. These two horizons merge into a single one (extreme black hole) when the model parameter satisfies $\tfrac{a}{M}=\tfrac{4\sqrt{3}}{9}$, whereas for larger values no horizon arises, i.e., the wormhole throat uncloaks and one finds a family of traversable wormholes. The behaviour of the metric function $A(x)$ is depicted in Fig. \ref{fig:metric} for different representative cases of these configurations, and which shall be further used later when discussing the geodesic motion and shadows.

For the purpose of building the main equations for geodesic motion, we recall that a photon follows a null geodesic of the background metric, $g^{\mu\nu}k_{\mu}k_{\nu}=0$, where $k_{\mu}=\dot{x}_{\mu}$ (dots denote derivatives with respect to an affine parameter) is the tangent vector to the photon trajectory. For static space-times, and assuming $\theta=\pi/2$  by spherical symmetry without loss of generality, there are two conserved quantities, namely, the energy per unit mass, $E=A(x)\dot{t}$, and the angular momentum per unit mass, $L=r^2(x)\dot{\phi}$, which for the sake of the geodesic motion are combined into a single {\it impact parameter} $b \equiv L/E$. This turns the null geodesic equation into
\begin{equation} \label{eq:geoeq}
\dot{x}^2=\frac{1}{b^2}-V_{eff}(x) \ ,
\end{equation}
where the effective potential reads off as
\begin{equation} \label{eq:Veff}
V_{eff}(x)=\frac{A(x)}{r^2(x)} \ .
\end{equation}
The geodesic equation (\ref{eq:geoeq}) determines the fate of a given light ray depending on its impact parameter. In particular, when at some $x=x_0$ the impact parameter of a photon fulfils the equation $1/b^2=V_{eff}(x_0)$, then the vanishing of the right-hand side of (\ref{eq:geoeq}) implies that the photon is deflected at the minimum distance $x_0$ from the compact object. The smallest (critical) impact parameter for which this may happen corresponds to the one for which Eq.\eqref{eq:geoeq} vanishes at the maximum of the potential, i.e. (here $'\equiv d/dx$)
\begin{equation} \label{eq:critcurve}
b_c=V^{-1/2}(x_c), V'(x_c)=0, V''(x_c)<0 \ ,
\end{equation}
where the corresponding radius $x_c$ is determined by the solution of the equation $(r^2)'/r^2=A'/A$. Geodesic trajectories with the critical impact parameter defined in Eq.\eqref{eq:critcurve} are known as critical curves or simply as photon spheres. Any such a light ray will undergo an arbitrarily large number of orbits around its critical curve until a small radial perturbation will cause it to either run away free to asymptotic infinity or to be dragged by the compact object and eventually hit the event horizon (in the black hole cases) or the throat (in the traversable wormhole case).

\begin{figure}[t!]
\includegraphics[width=8.0cm,height=5.5cm]{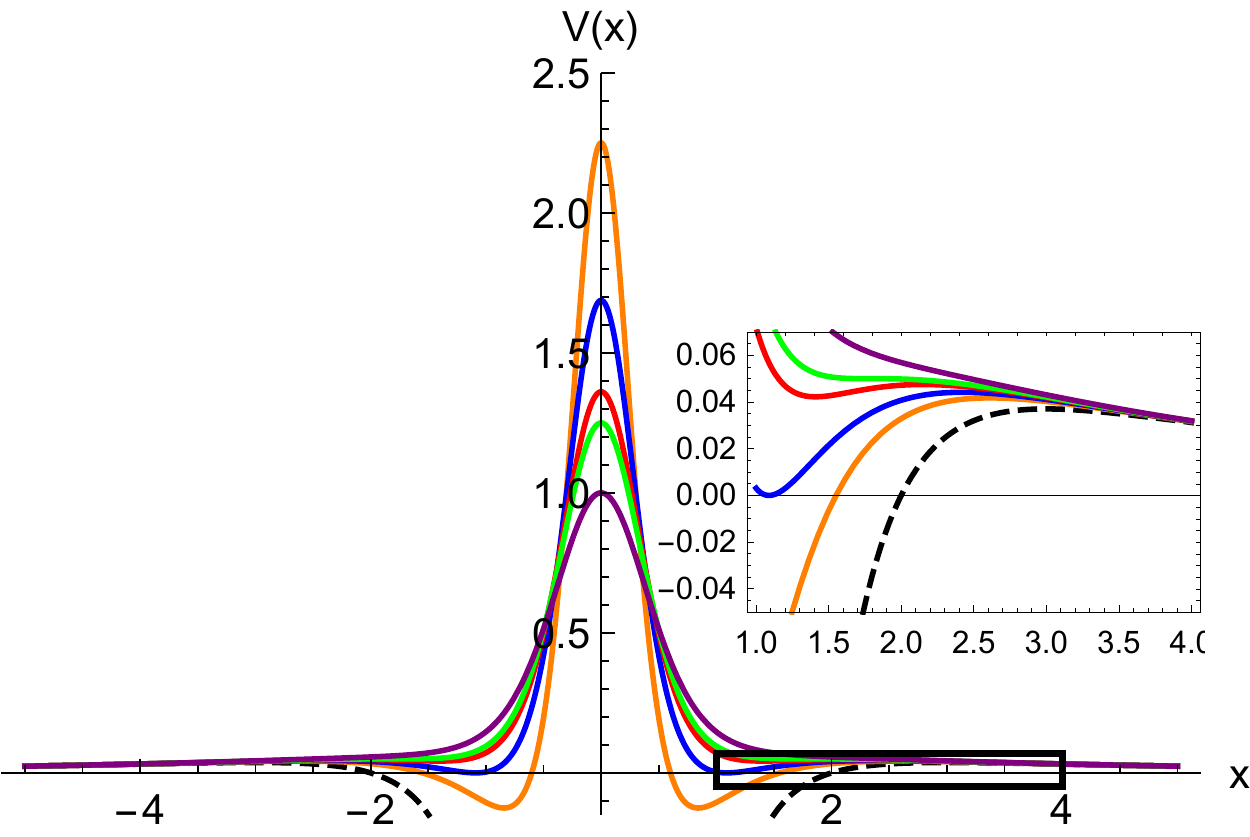}
\caption{The effective potential $V(x)$ (in units of $M=1$) for the configurations with $a=0$ (dashed black), $a=2/3$ (orange), $a=\tfrac{4\sqrt{3}}{9}$ (blue), $a=6/7$ (red), $a=\tfrac{2\sqrt{5}}{5}$ (green) and $a=1$ (purple). For non-vanishing $a$ the absolute maximum of the potential is located at $x=0$, while for $a \leq \tfrac{2\sqrt{5}}{5}$ another (local) maximum is also present [see the inset figure].}
\label{fig:potential}
\end{figure}

In Fig. \ref{fig:potential}, we depict the effective potential for several relevant values of the parameter $a/M$. As it can be seen there, any non-vanishing value of $a/M$ induces strong changes in the shape of the effective potential as compared with the one of the Schwarzschild solution. Indeed, the most salient feature of this family of solutions is that the potential attains its absolute maximum value at $x=0$, which effectively acts as a photon sphere at some (model-dependent) impact parameter $b_c^1$. This photon sphere, however, is hidden behind an event horizon if $\tfrac{a}{M}<\tfrac{4\sqrt{3}}{9}$, preventing any light ray with origin below the event horizon to reach an asymptotic observer. On the other hand, a second impact parameter $b_c^2$ is present as associated to a (local) maximum in this potential provided that $\frac{a}{M} \leq \tfrac{2\sqrt{5}}{5}$, which is always accessible since the event horizon (when present) lies below.

Therefore, this family describes the following configurations relevant for the analysis of their optical appearances, depending on the parameter $a/M$:

\begin{itemize}

\item $a=0$: Schwarzshild black hole, where the throat closes since $x^2 \to r^2$. A single photon sphere is present located at  $x_c^{(2)}=3M$, reached by photons with an impact parameter $b_c^2=3\sqrt{3}M$.
\item $0<\tfrac{a}{M}<\tfrac{4\sqrt{3}}{9}$: a black hole with two horizons having a single accessible photon sphere at some $x_c^{(2)}$ with an impact parameter $b_c^2$.
\item $\tfrac{a}{M}=\tfrac{4\sqrt{3}}{9}$: an extremal black hole, since the two horizons from the previous case merge into a single one. Nevertheless, just the outer photon sphere is accessible.
\item $\tfrac{4\sqrt{3}}{9}<\tfrac{a}{M}<\tfrac{2\sqrt{5}}{5}$: a traversable wormhole having two photon spheres reached by impact parameters $b_c^1$ and $b_c^2$ respectively. In the intermediate region, $b_c^1<b<b_c^2$, the minimum of the potential acts as an anti-photon sphere (a stable bound orbit\footnote{It should be stressed that ultra-compact objects having an anti-photon sphere may allow for trapped long-lived modes, therefore raising the issue of whether they can grow large enough to destabilize the whole system \cite{Cardoso:2014sna}, which requires an analysis of the stability of any such objects on a case-by-case basis.}).
\item $\tfrac{a}{M}=\tfrac{2\sqrt{5}}{5}$: a traversable wormhole with one photon sphere corresponding to the maximum of the potential at $x_c=0$. The first derivative of the potential vanishes also at another (outer) point, which corresponds to an inflection point and consequently not to a bounded orbit.
\item $\tfrac{a}{M}>\tfrac{2\sqrt{5}}{5}$: a traversable wormhole with a single photon sphere at $x_c=0$ with impact parameter $b_c^1$.
\end{itemize}

\begin{figure}[t!]
\includegraphics[width=8.0cm,height=5.5cm]{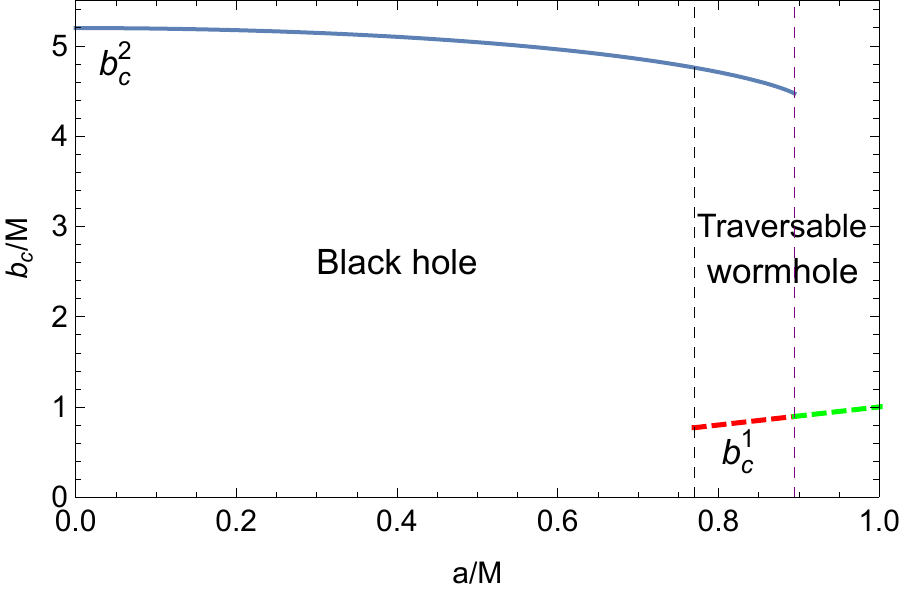}
\caption{The critical impact parameter $b_c/M$ as a function of $a/M$ for the outer curve ($b_c^2$, solid blue) and the inner one ($b_c^1$, dashed) within the range $a/M \in[0,1]$. In the latter we have depicted in red the case in which both critical curves exist for the same configuration, and in green when only $b_c^1$ is present.}
\label{fig:critical}
\end{figure}

The impact parameter for the cases with an inner critical curve is simply given by $b_c^1=a$, while for the outer critical curve the impact parameter, $b_c^2$, has quite a cumbersome expression as a function of $a/M$, which is numerically depicted in Fig. \ref{fig:critical}. For the sake of this work we shall take several choices of parameters representing the different configurations described above and summarized in Table \ref{table:I}, alongside the specific values of the associated impact parameters and their locations in the $x$-coordinate.  In the sequel, we shall actually be interested in comparing a sample of a two-horizons black hole and a wormhole configuration in order to compare their respective shadows and lights rings on equal footing.

\begin{table}[]
\begin{tabular}{|c|c|c|c|c|c|c|}
\hline
$a/M$ & $0$  & $2/3$  & $4\sqrt{3}/9$  & $6/7$  & $2\sqrt{5}/5$  &  $1$ \\ \hline
Type  & Sch  & 2hBH  & EBH  & TWH  & TWH  & TWH  \\ \hline
$b_c^1/M$ & X & X & X   & $6/7$  & $2\sqrt{5}/5$  & $1$  \\ \hline
$b_c^2/M$ & $5.1961$  & $4.8936$ & $4.7585$   & $4.5888$  & $4.4721$  & X  \\ \hline
$x_m^{(2) \vert (1)}$ & $3$ & $2.6099$ & $2.4218$  & $2.1493 \vert 0$  & $1.7888 \vert 0$  & $0$  \\ \hline
$b_{is}$ & $2.8519$ & $2.3922$ & $1.88187$   & $0.7201$  & $0.7497$  & $0.8336$   \\ \hline
\end{tabular}
\caption{The features of several chosen configurations depending on the parameter $a/M\in[0,1]$. In this table we depict the type of configuration (Schwarzschild, 2-horizons black hole, extreme black hole, and traversable wormhole), the critical parameter (in units of $M=1$) for the inner, $b_c^1$, and outer, $b_c^2$, photon spheres (when they exist) and their associated location in the coordinate $x$. Furthermore, $b_{is}$ denotes the impact parameter for the inner shadow, below which no intersections with the equatorial plane are made (see Sec. \ref{sec:Ray} for details).}
 \label{table:I}
\end{table}

\section{Ray-tracing} \label{sec:Ray}

In order to determine the optical appearance of a compact object one needs to integrate the geodesic equation (\ref{eq:geoeq}), conveniently rewritten as the variation of the azimuthal angle with respect to the radial coordinate $x$ as
\begin{equation} \label{eq:geoeq2}
\frac{d\phi}{dx}= \mp \frac{b}{r^2(x)\sqrt{1- \tfrac{b^2A(x)}{r^2(x)}}} \ ,
\end{equation}
where $\mp$ for ingoing(outgoing) geodesics. Since the image on the observer's screen is the gravitationally lensed trajectories of light rays coming from different parts of the sky, it is convenient to employ the ray-tracing method. In this method, the trajectory of each light ray arriving to the observer's screen is backtracked using Eq.(\ref{eq:geoeq2}) to determine the point of the sky where it was originated from. Depending on the impact parameter of a particular light ray, this might turn around the compact object a certain number of times on its path for $b>b_c$, allowing to divide the impact parameter region according to the (normalized) number of (half-)orbits, $n \equiv \tfrac{\phi}{2\pi}$, which  accounts for the number of intersections with the equatorial plane of a particular right ray, $m=[2n]$.

We conveniently orientate our setup in order to locate the observer on the (far) right-hand side of the screen (corresponding to the north pole). Therefore, in this language, a photon not being deflected at all by the compact object (traveling on a straight line from left to right on this plane) has $n=1/2$. As we get close to the critical impact parameter, $b \gtrsim b_c$, a light ray will increase its number of orbits until formally going to infinity at the critical value. Since in our subsequent analysis the accretion disk surrounding the compact object will be assumed to be optically thin, on each intersection with the equatorial plane every light ray will pick up additional brightness in a way that largely depends on the particular assumed emission modelling of the disk (see Sec. \ref{Sec:accdisk}). In the geometrically thin disk setting, this produces a infinite sequence of concentric rings from photons that have completed $n$ half-orbits in their approach to the critical curve. However, since the impact parameter window allowing a certain number of half-orbits is quickly diminished, the contribution of successive orbits to the total luminosity is generically expected to be strongly suppressed (see \cite{Bisnovatyi-Kogan:2022ujt} for a general discussion). From a practical point of view, therefore, the relevant contributions to the total luminosity on the observer's screen will be provided by three types of trajectories:
\begin{itemize}
\item Direct emission: corresponding to light rays intersecting the equatorial plane (on its front) just once, and defined by $1/2<n\leq 3/4$ ($m=1$). This is the dominant contribution to the optical appearance of the object, both in terms of luminosity and width of the associated ring of radiation.
\item Lensed emission: corresponding to light rays intersecting the equatorial plane twice (on its front and its back, respectively), and defined by $3/4<n \leq 5/4$ ($m=2$), being the sub-dominant contribution to the luminosity.
\item Photon ring emission:  corresponding to light rays intersecting the equatorial plane at least thrice, and defined by $n>5/4$ ($m=3$).
\end{itemize}
After three intersections with the equatorial plane, $m>3$, the corresponding light rings are expected to be so demagnified \cite{Gralla:2019drh} that their contribution to the total luminosity can be dismissed\footnote{In Ref.\cite{Vincent:2020dij} the authors use the alternative notion of {\it secondary rings} as those trajectories approaching the critical curve and that visit those regions of the accretion disk emitting most of the radiation, so as to be non-negligible from the point of view of the luminosity on the observer's screen. For the sake of this paper, this notion is fully equivalent to the two types of trajectories beyond the direct one.}. Therefore, we shall reserve the word {\it photon ring} for the closer ring to the critical curve.

On the other hand, for impact parameters $b<b_c$ the light ray will also perform a number of half-orbits on its trip down to intersect the event horizon (in the black hole cases) or the throat (in the wormhole cases). As we shall see in Sec. \ref{Sec:accdisk}, in accretion disk models where the inner edge of the disk is allowed to extend inside the outer photon sphere, part of that emission will be able to leak off until reaching the observer's screen. Thus, one can also establish a similar classification of the different trajectories in terms of direct/lensed/photon ring ones according to the number of half-orbits (i.e. intersections with the equatorial plane). However, in this case the range of the direct trajectories is extended to  $1/4 \leq n \leq 3/4$, where the lower limit is defined by the requirement that there is at least one intersection with the equatorial plane (note that in this $b<b_c$ region, straight motion corresponds to $n=0$). The region below of this lower limit defines the inner shadow \cite{Chael:2021rjo} which, being a feature depending only on the background geometry, is the absolute blackness region (in the black hole case) no matter the details of the accretion disk.

\begin{figure}[t!]
\includegraphics[width=8.0cm,height=5.5cm]{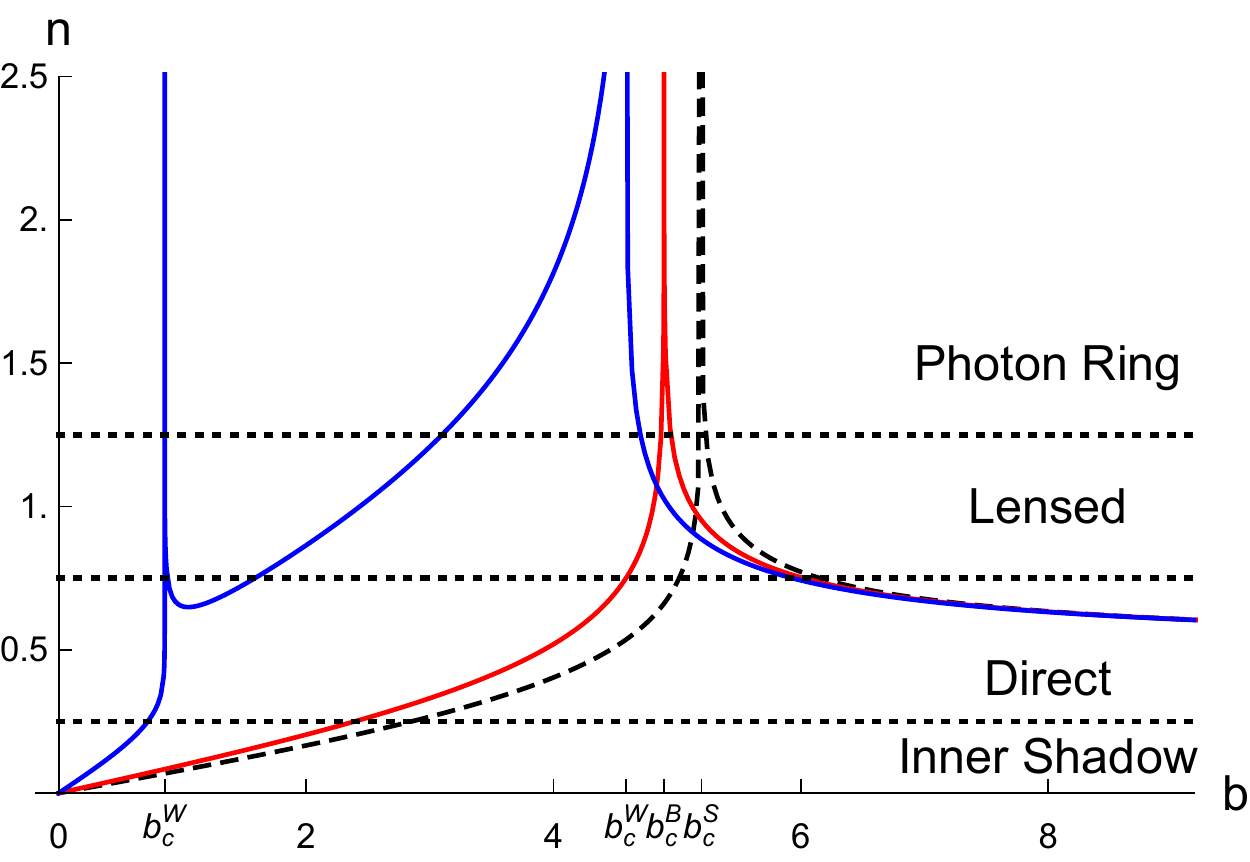}
\caption{The number of half-orbits $n \equiv \phi/2\pi$ as a function of $b$ (in units of $M=1$) for the Schwarzschild black hole ($a=0$, dashed black), two-horizons black hole ($a=2/3$, red) and traversable wormhole case having two photon spheres ($a=6/7$, blue). The direct, lensed, and photon ring emission regions are enclosed by horizontal dotted lines, and we also depict the inner shadow limit. In this plot $b_c^{S}$, $b_c^{B}$ and $b_c^{W}$ (two) are the critical impact parameters for these three configurations respectively.}
\label{fig:orbits}
\end{figure}

In Fig. \ref{fig:orbits} we depict the number of half-orbits for the two chosen samples of two-horizons black hole, $a=2/3$, and traversable wormhole having two photon spheres, $a=6/7$, as compared to the Schwarzschild black hole of GR, $a=0$. The location of the critical curve on each case is marked by the divergence in the number of half-orbits. In the black hole cases, the single critical curve takes the form of a spike, while in the traversable wormhole case one observes the two divergences corresponding to the two photon spheres, but while the inner one looks like a very sharp spike the outer one is much more wider. Note that in the intermediate region between such two critical curves the number of half-orbits remains mostly in the lensed/photon ring regions.  This plot hints at the features one can expect from the ray-tracing of a bunch of geodesics in both the black hole and the wormhole cases, which we study separately.

\begin{figure*}[t!]
\includegraphics[width=8.5cm,height=8.0cm]{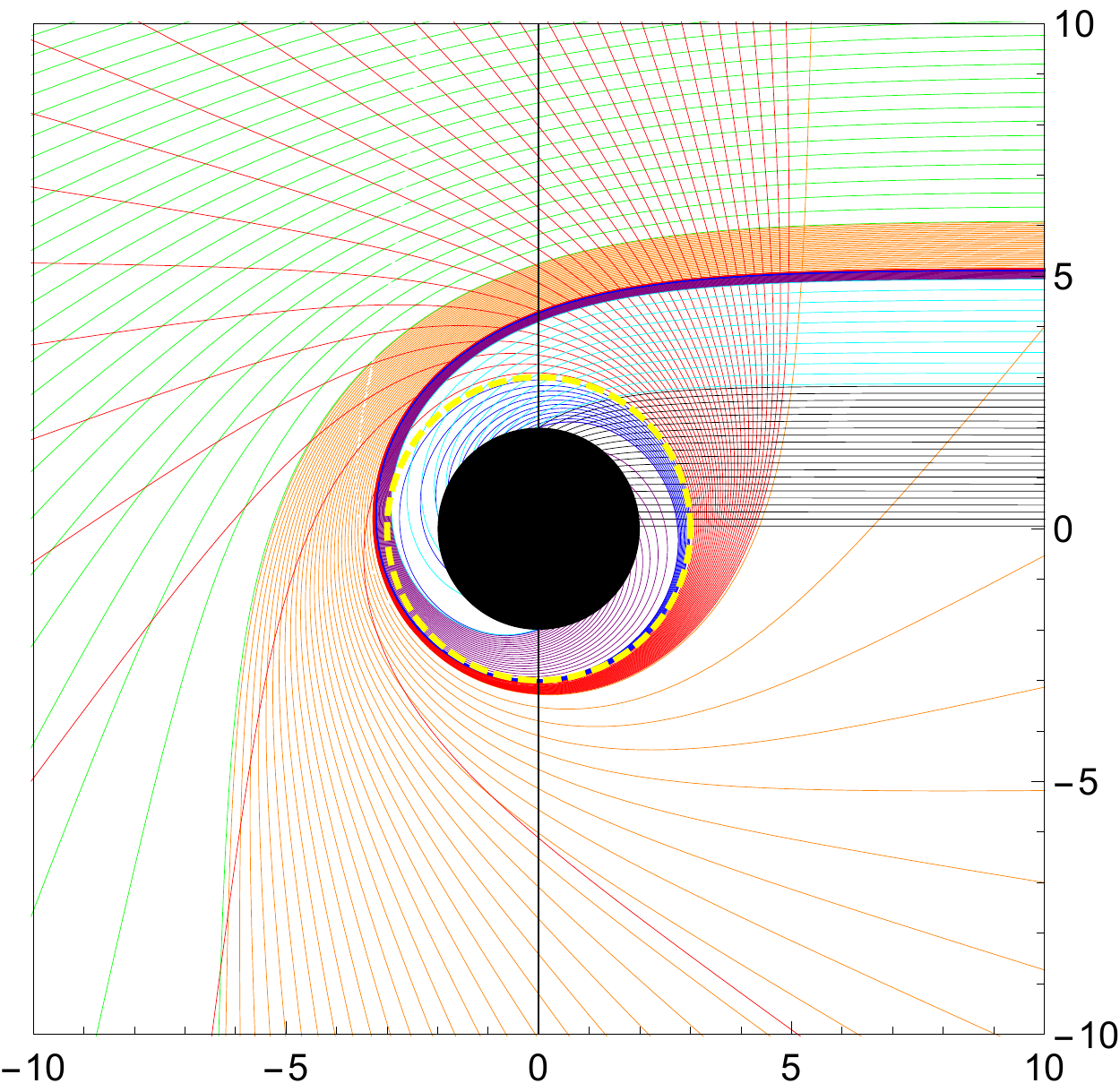}
\includegraphics[width=8.5cm,height=8.0cm]{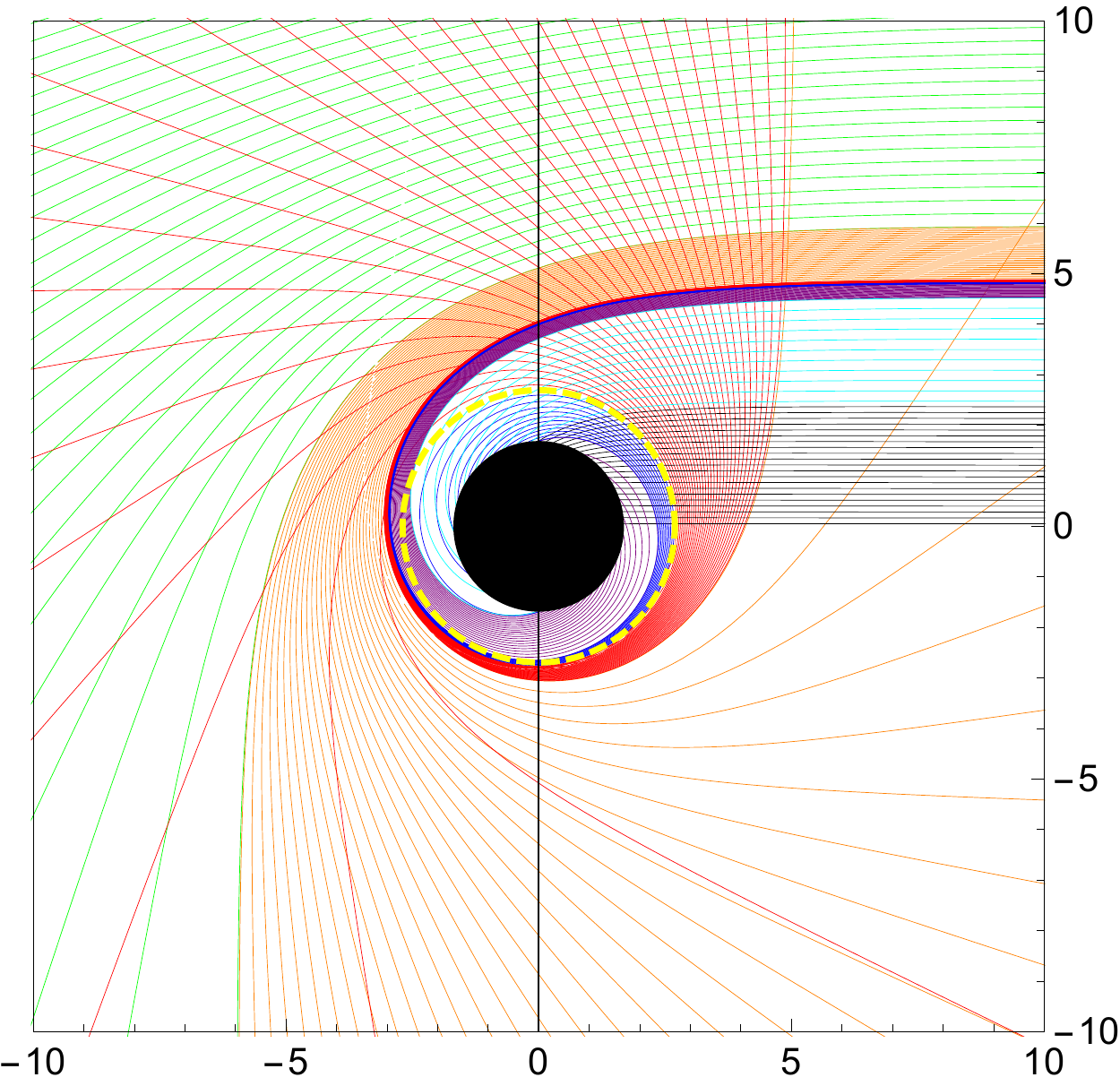}
\caption{Ray-tracing of the black hole configurations (in units of $M=1$) with $a=0$ (Schwarzschild, left) and $a=2/3$ (two-horizons black hole, right) for a range of relevant values of the impact factor, making use of the radial function $r$.  The observer's screen is located in the far right side of this plot and the type of emission is defined with respect to the number of intersections with the equatorial plane (vertical line): for $b>b_c$ we have direct (green), lensed (orange) and photon ring (red) emissions reaching to a minimum distance from the photon sphere (dashed yellow circumference) before running away, while for $b<b_c$ we also have direct (cyan), lensed (purple) and photon ring (blue) emissions. The latter three trajectories intersect the black hole horizon (black central circle) after crossing the photon  sphere. The bunch of black curves do not intersect the equatorial plane and therefore no emission can come out on them no matter the accretion disk model, therefore corresponding to the inner shadow of the solutions.}
\label{fig:raytracBH}
\end{figure*}

\subsection{Black hole with two horizons}

Let us consider first the two-horizons black hole solution with $a=2/3$, whose properties are characterized in Table \ref{table:I}. The ray-tracing of this solution is depicted in Fig. \ref{fig:raytracBH} (right) as compared to the behaviour of the Schwarzschild solution of GR (left). While the corresponding plots for these two solutions are qualitatively similar, there are some differences in the widths of their impact parameter regions:
\begin{itemize}
\item Direct: $b/M >6.02$ and $b/M \in (2.39,4.59)$.

[Schwarzschild: $b/M >6.15$ and $b/M \in (2.85,5.02)$].

\item Lensed: $b/M \in (4.59,4.87)$ and $b/M \in (4.95,6.02)$.

[Schwarzschild: $\in (5.02,5.19)$ and $b/M \in (5.23,6.15)]$.

\item Photon Ring: $b/M \in (4.87,4.95)$.

 [Schwarzschild: $b/M \in (5.19,5.23)$].

\item Inner Shadow: $b/M<2.39$.

 [Schwarzschild: $b/M<2.85$].
\end{itemize}
Therefore, one can see a moderate decrease in the contribution of the direct emission to the impact parameter region as compared to the lensed/photon ring ones, and a shrinking in the maximum value of the impact parameter defining the beginning/ending of a given type of emission, which is consistent with the analysis of the number of half-orbits depicted in Fig. \ref{fig:orbits}. This effect is naturally induced by the diminished photon sphere radius and black hole horizon as compared to the Schwarzschild black hole of GR.

\begin{figure*}[t!]
\includegraphics[width=8.5cm,height=8.0cm]{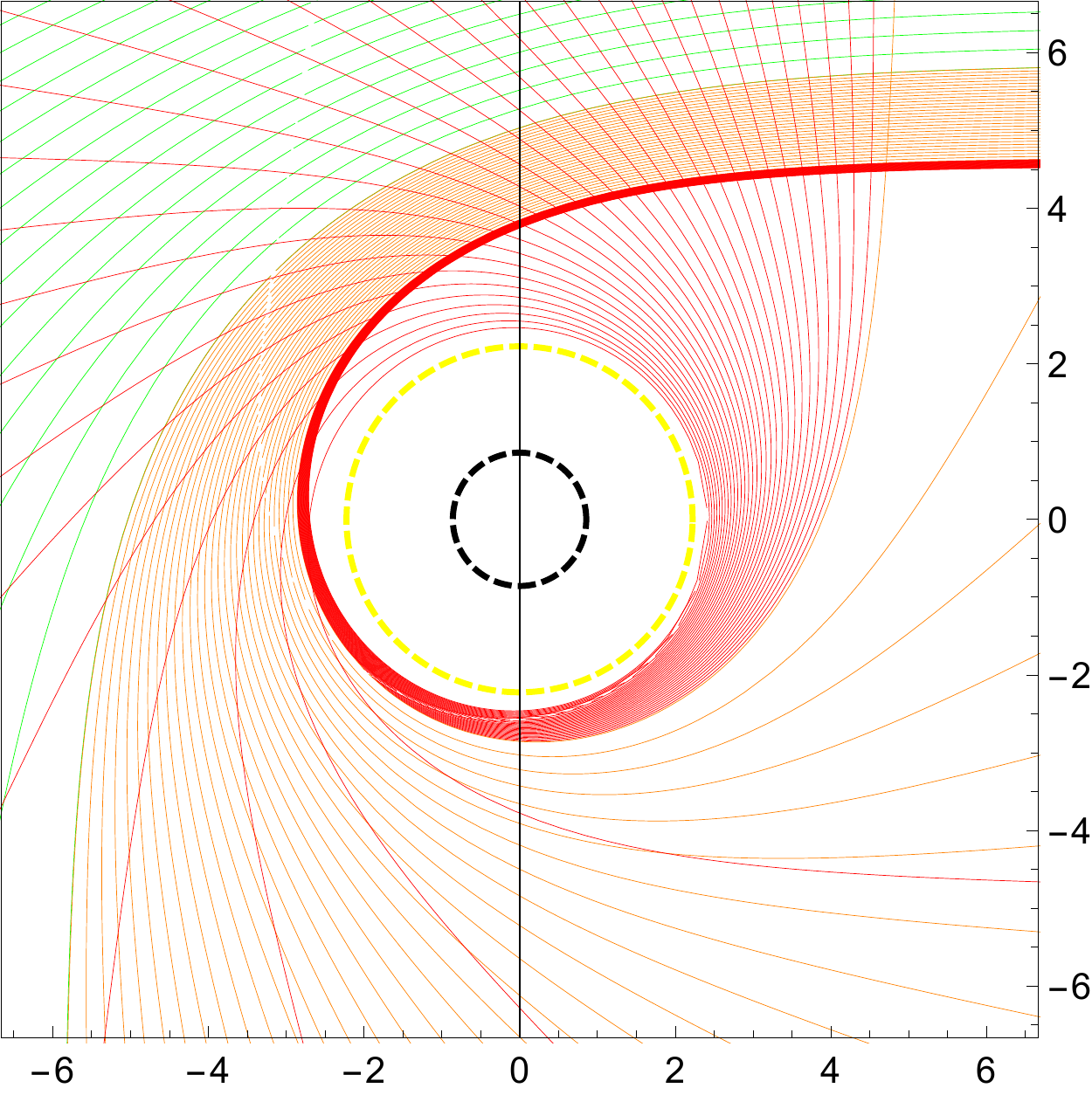}
\includegraphics[width=8.5cm,height=8.0cm]{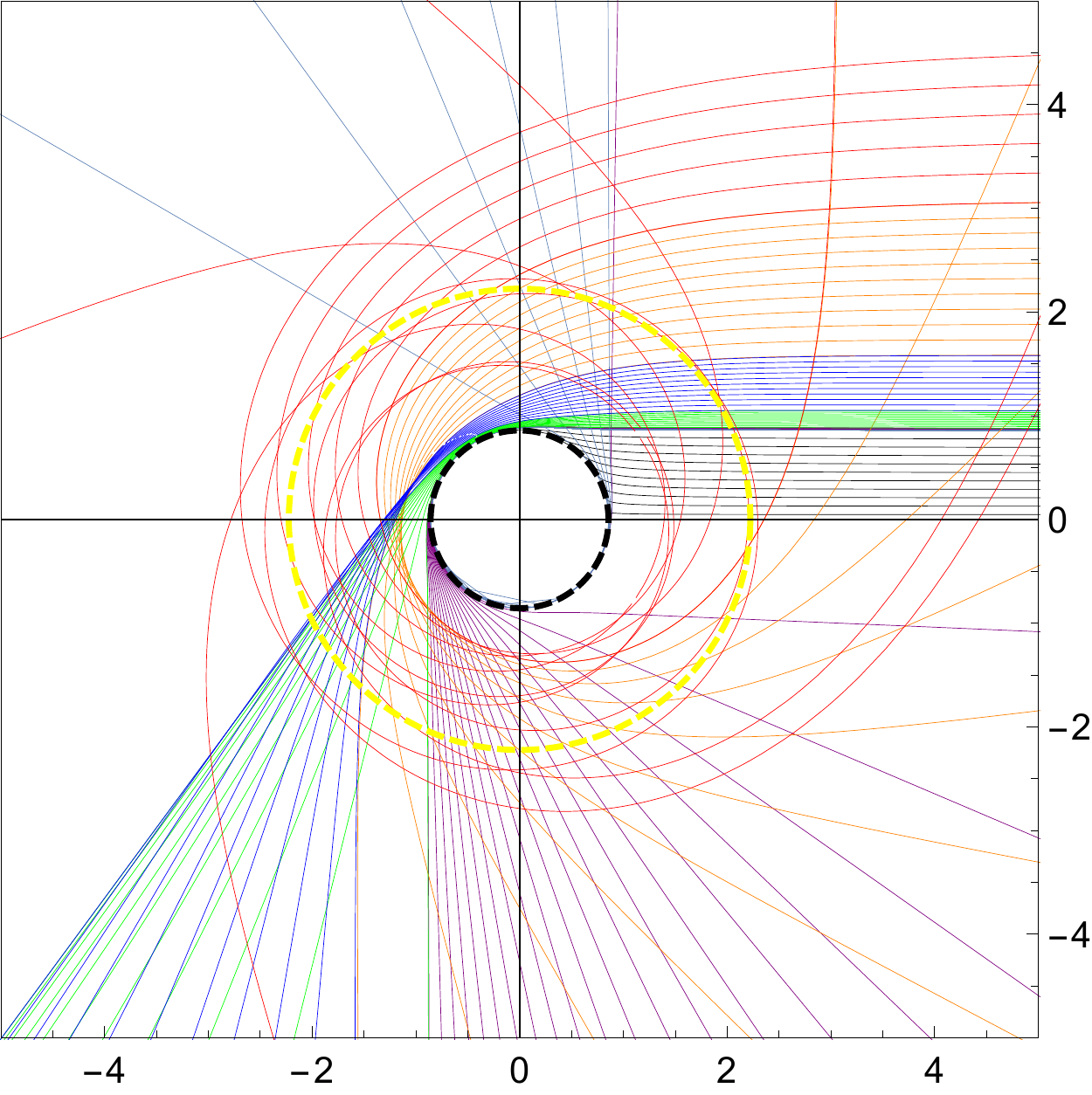}
\caption{Ray-tracing of the traversable wormhole configurations having two photon spheres with (in units of $M=1$) $a=6/7$, for impact parameters above (left plot) and below (right plot) the outer critical one $b_c^2$ (dashed yellow  circumference). On the left plot, $b>b_c^2$, direct (green), lensed (orange) and photon ring (red) trajectories follow the same logic as in the black hole case of Fig. \ref{fig:raytracBH}. On the right plot, we find for $b_c^1<b<b_c^2$ the outer photon ring (red), lensed (orange), and direct (blue) trajectories, as well as the inner direct (green), lensed (purple) and photon ring (cyan) trajectories.  Below $b<b_c^1$ we also find the inner (blank rather than black) shadow, which is only potentially accessible to those light rays travelling from the other side of the wormhole. }
\label{fig:raytracWH}
\end{figure*}

\subsection{Traversable wormhole having two photon spheres}

Things get far more interesting in the traversable wormhole solutions having two photon spheres, whose ray-tracing for the model parameter $a/M=6/7$ is depicted in Fig. \ref{fig:raytracWH} for trajectories above the outer critical impact parameter (left panel) and below of it (right panel). While those light rays with $b>b_c^2$ follow a similar pattern as those of the black hole case discussed above, having direct ($b/M>5.93$), lensed ($b/M \in (4.70,5.93)$) and photon ring ($b/M \in (4.59,4.70)$) contributions, those in the region between the two photon spheres, $0.85714M \approx b_c^1<b<b_c^2 \approx 4.5888M$, have a frenzied behaviour. This is particularly true for those light rays which hover just slightly below the critical impact parameter $b \lesssim b_c^2$, having quite a chaotic pattern of trajectories between the two photon spheres before being finally able to exit the outer one. Decreasing further the impact parameter the trajectories have a more stable pattern and one finds the following trajectories driven by the combined influence of the outer and inner critical curves:
\begin{itemize}
\item Inner-outer photon ring: $b/M \in (3.10,4.59)$.
\item Inner-outer lensed: $b/M \in (1.59,3.10)$.
\item Inner-outer direct: $b/M \in (0.8796,1.59)$.
\item Inner-outer lensed: $b/M \in (0.8572,0.8796)$.
\item Inner photon ring: $b/M \in (0.857143,0.85720)$.
\item Inner lensed: $b/M \in (0.857139,0.857143)$.
\item Inner direct: $b/M \in (0.7201,0.857139)$.
\item Inner (blank) shadow: $b/M<0.7201$.
\end{itemize}
where we are characterizing the direct/lensed/photon ring trajectory on each case depending on whether they originated in the valley between the inner and outer critical curves, or at the inner curve itself. In particular, the trajectories at the bottom of the valley  show a non-monotonic behaviour in the sense that the corresponding bunch of light rays intersect each other near their respective boundaries, due to the minimum attained in the number of half-orbits in the transition from the outer to the inner critical curve (recall Fig. \ref{fig:orbits}).

As it can be seen from this discussion and its associated plot, the range for the outer lensed/photon ring emissions is largely enhanced as compared to the black hole case due to the complex interplay between the two photon spheres. Nonetheless, and as it was expected, the impact parameter region for the lensing/photon ring emissions  driven by the inner critical curve is extremely narrow: due to this fact, for the sake of the emission from accretion disks of Sec. \ref{Sec:accdisk} we shall extend the inner region to the photon ring trajectories above to include the full region $b_{is}<b<b_c^1$. The inner shadow is named here as blank since light rays coming from the other side might flow through the wormhole throat and reach eventually the observer, colouring the central darkness region of the black hole case (and also reaching significantly smaller impact factor values), though the detailed analysis of such a scenario goes beyond the scope of this work.

\section{Shadows from geometrically thin accretion disks} \label{Sec:accdisk}

Though in the previous section we have modelled the shadow of the black hole/wormhole as the central depression of the image as seen by a far-away observer, and the shapes of the light rings in terms of the number of half-orbits, these are highly idealized observables, while realistic astrophysical images are mainly fuelled by the physics of the accretion disk around the compact object. Though a precise modelling of this problem requires the use of general relativistic magneto-hydrodynamic (GRMHD) simulations, significant progress can also be made on the theoretical front by using analytical models of static accretion disks with a localized emission on a given geodesic starting from a finite-size region of the disk.

\begin{figure*}[t!]
\includegraphics[width=5.9cm,height=5.0cm]{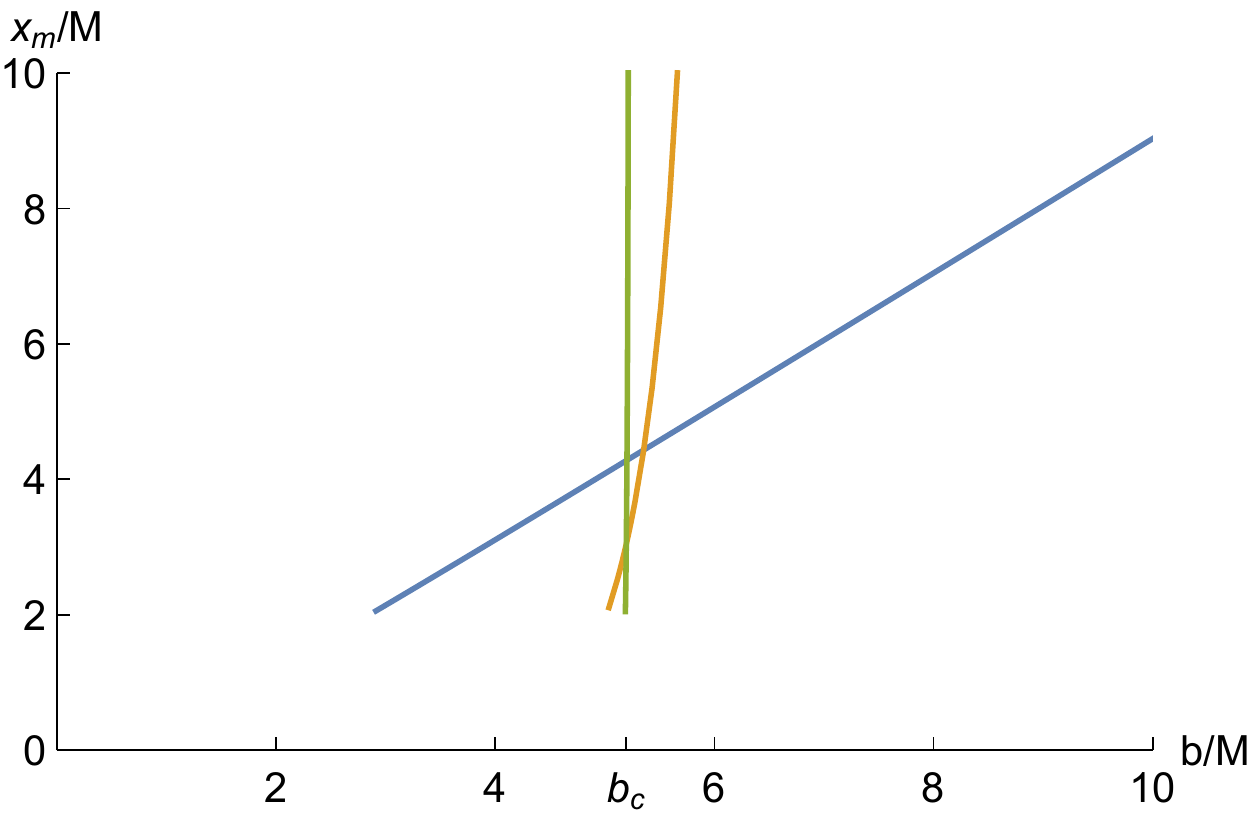}
\includegraphics[width=5.9cm,height=5.0cm]{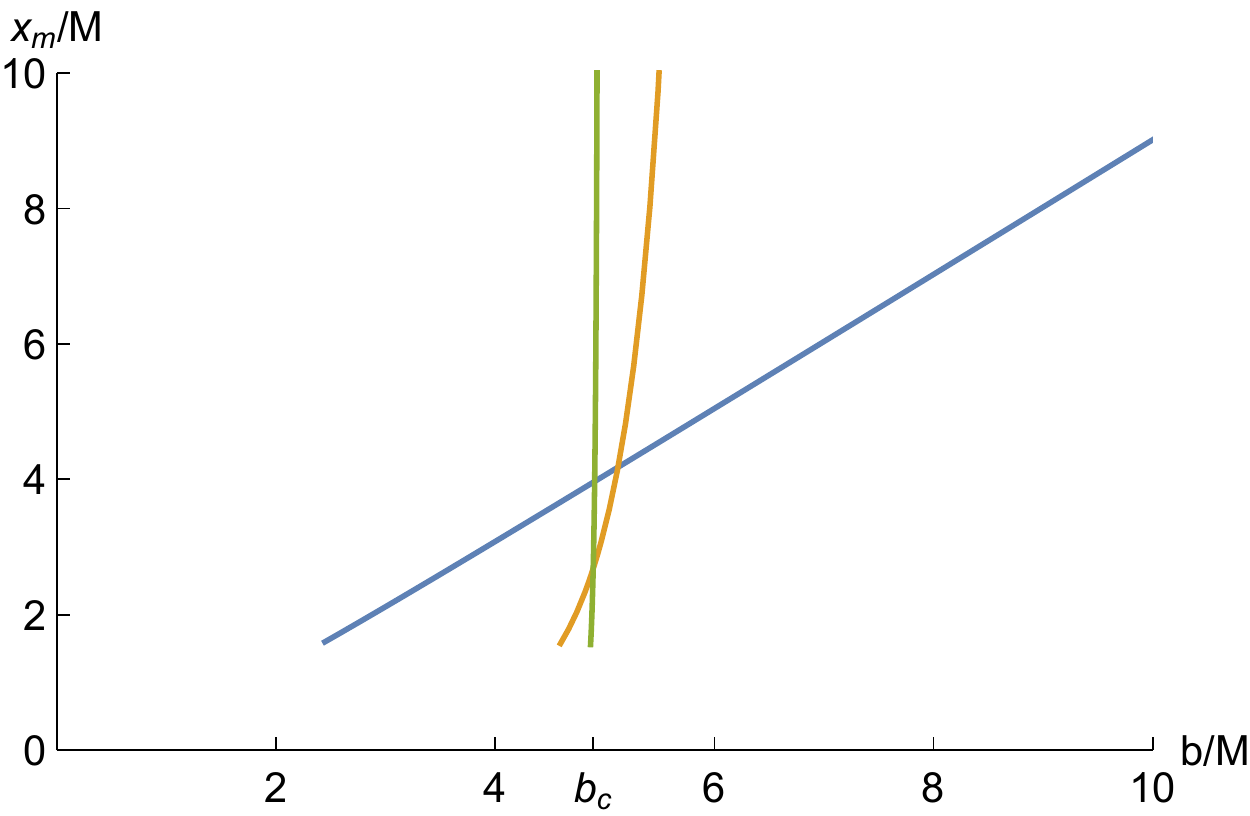}
\includegraphics[width=5.9cm,height=5.0cm]{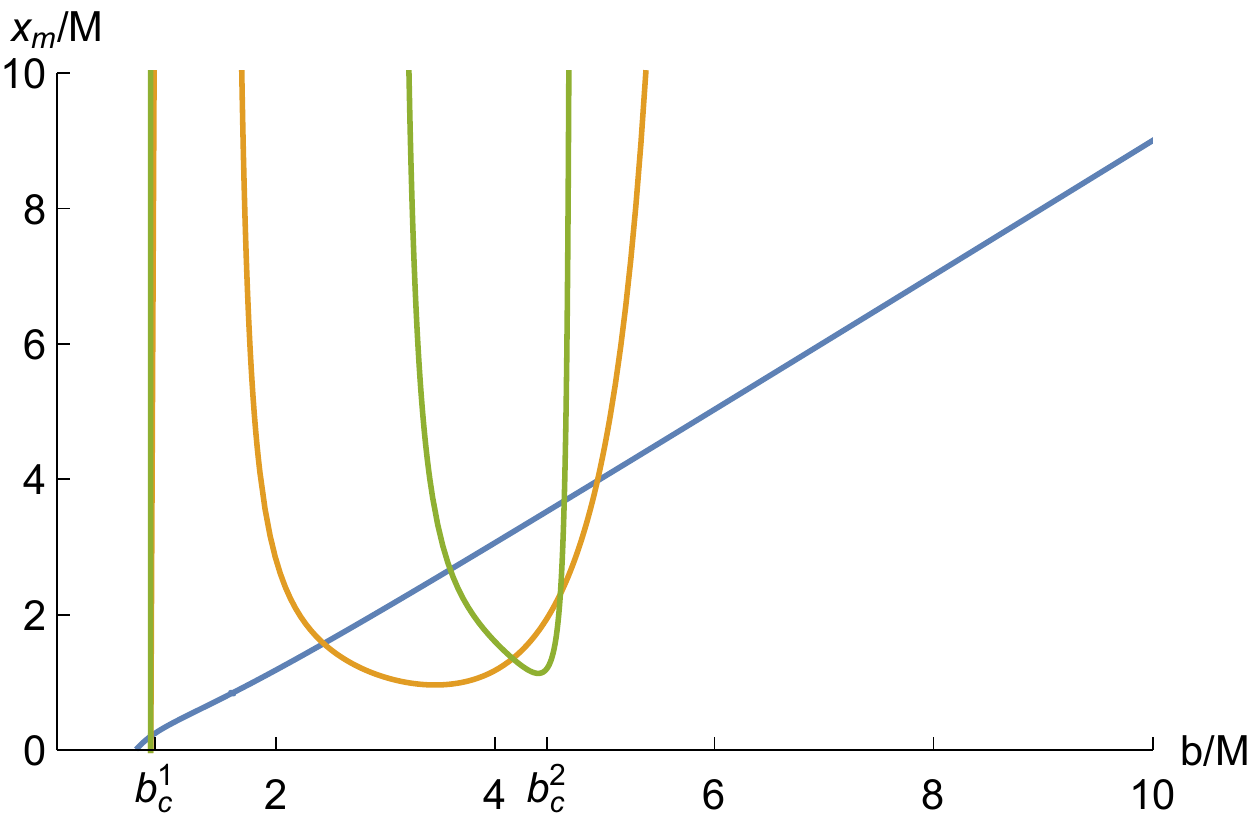}
\caption{The first three transfer functions for the direct (blue), lensed (orange) and photon ring (green) emissions for the Schwarzschild solution (left), black hole with $a/M=2/3$ (middle), and traversable wormhole with $a/M=6/7$ (right). $b_c$ denotes the location of the corresponding photon sphere(s), with the black hole cases having the standard single one, while in the traversable wormhole case a second one is present. The slope of each curve is interpreted as the demagnification factor of the corresponding emission.}
\label{fig:transfunc}
\end{figure*}

In order to model the emission from a finite-size disk one starts (neglecting scattering) from the radiative transfer (Boltzmann) equation, which is written as \cite{RadiativeBook}
\begin{equation} \label{eq:Bol}
\frac{d}{d \lambda}\left(\frac{dI_{\nu}}{d\nu^3}\right)=\left(\frac{j_\nu}{\nu^2}\right)  -(\nu \alpha_\nu)\left(\frac{I_\nu}{\nu^3}\right) \ ,
\end{equation}
where $I_{\nu}$ is the intensity for a given frequency $\nu$, $j_{\nu}$ is the emissivity, $\alpha_\nu$ the absorptivity, and quantities inside parenthesis are frame-independent. The resolution of the above equation requires feeding it with precise knowledge of the plasma fluid (e.g. number density, angular momentum, emissivity and absorptivity) making up the disk, to be implemented in GRMHD simulations. Nonetheless, the results of such simulations \cite{Gold:2020iql} hint that analytical models can be implemented by first neglecting absorption effects, $\alpha_\nu =0$, and next by assuming a source which emits monochromatically, $j_\nu \sim \nu^2$. Moreover, for a geometrically (infinitesimally) thin accretion disk (located in the vertical line of the ray-tracing plots), Eq.(\ref{eq:Bol}) implies that $I_{\nu}/\nu^3$ is conserved along a photon's trajectory. Furthermore, we shall assume an isotropic emission, i.e., $I^{em}_{\nu}=I(x)$. For the purpose of simulating different stages in the temporal evolution of such an accretion disk we shall employ three canonical toy models whose inner edge (assumed to represent the effective source of emission of the disk) extends up to some relevant surface, while  smoothly falling off asymptotically with different tails. Specifically, such models are defined as follows:

\begin{itemize}

\item Model I: It starts its emission at the innermost stable circular orbit for time-like observers (ISCO), modelled as
\begin{equation}
I^{em}_I(x)=\frac{1}{(x-(x_{isco}-1)^2}
\end{equation}
if $x>x_{isco}$ and zero otherwise.

\item Model II: It starts its emission at the outer critical curve, modelled as
\begin{equation}
I^{em}_{II}(x) =\frac{1}{(x-(x_{ps}-1))^3}
\end{equation}
if $x>x_{ps}$ and zero otherwise.

\item Model III: Its emission goes all the way down to the horizon (in the black hole case\footnote{From the point of view of the GRMHD simulations relevant for the EHT observations this is the most suitable scenario \cite{Gold:2020iql}. }) or to the throat (in the wormhole case), modelled as
\begin{equation}
I^{em}_{III}(x) =\frac{\pi/2-\arctan[x-5]}{\pi/2-\arctan[\tilde{x}-5]}
\end{equation}
(where $\tilde{x}=x_{h}$ in the black hole case and $\tilde{x}=x_{th}=0$ in the wormhole case) and zero otherwise.

\end{itemize}

The intensity received on the observer's screen will be the emitted one corrected by two factors: firstly, it will be gravitationally redshifted in their winding off the compact object and, secondly, the additional intersections of the lensed and photon ring trajectories with the accretion disk will contribute to pick up additional luminosities according to the emission profile of the disk. To take into account the first effect, one notes that if the frequency of the photon in the rest frame of the gas in the disk is given by $\nu_e$ with associated intensity $I_{\nu_e}$, then, by Liouville's theorem, the photon frequency measured by the distant observer will be $\nu_o$ with  intensity $I^{ob}_{\nu_0}=(\nu_e/\nu_0)^3 I_{\nu_e}^{em}$. In the spherically symmetric geometry considered in this work this implies that $I^{ob}_{\nu_0}=A^{3/2}(x)I_{\nu_e}$. Integrating over the full spectra of frequencies, $I^{ob}=\int d\nu_e I^{ob}_{\nu_e}$, one finds the result $I^{ob}=A^2(x)I(x)$ \cite{Gralla:2019xty}. In order to include the second effect, we just need to compute
\begin{equation}
I^{ob}=\sum_{m} A^{2}(x)I(x)\left|_{x=x_m(b)}\right. \ ,
\end{equation}
where the so-called {\it transfer function} $x_m(b)$ encodes the location of the $m$-th intersection of the light ray with impact parameter $b$ with the disk. For the purpose of this work, $m=1,2,3$ denotes the direct, lensed and photon ring emission, neglecting additional intersections with the disk since they will presumably contribute much less to the total luminosity. Indeed, one can verify it via the slope of the transfer function, $dx/db$, since it is a measure of the degree of demagnification of the image \cite{Gralla:2019xty}. As it can be seen in Fig. \ref{fig:transfunc}, where we depict the transfer functions for the two samples of two-horizon black hole ($a=2/3$) and traversable wormhole having two photon spheres ($a=6/7$) solutions (as compared to the Schwarzschild solution of GR) such a slope is steeper in the photon ring emission than in the lensed one, and in both of them much more than in the direct one. This means that in the black hole cases the direct emission will be the largest contribution to the total luminosity by far, with the lensed and photon ring ones being highly demagnified. However, in the traversable wormhole case the presence of the second (inner) photon sphere not only largely enhances the window of impact parameters for the lensed and photon ring emissions, but also significantly reduces the slope of both curves in the intermediate region between the two critical curves, as shown in Fig.~\ref{fig:transfunc}. This will supposedly have a larger impact in the total share of the observed luminosity budget of these contributions as compared to the direct one, as we shall see at once. Let us again analyze the black hole and wormhole cases separately.

\subsection{The two-horizons black hole}

\begin{figure*}[t!]
\includegraphics[width=5.9cm,height=4.8cm]{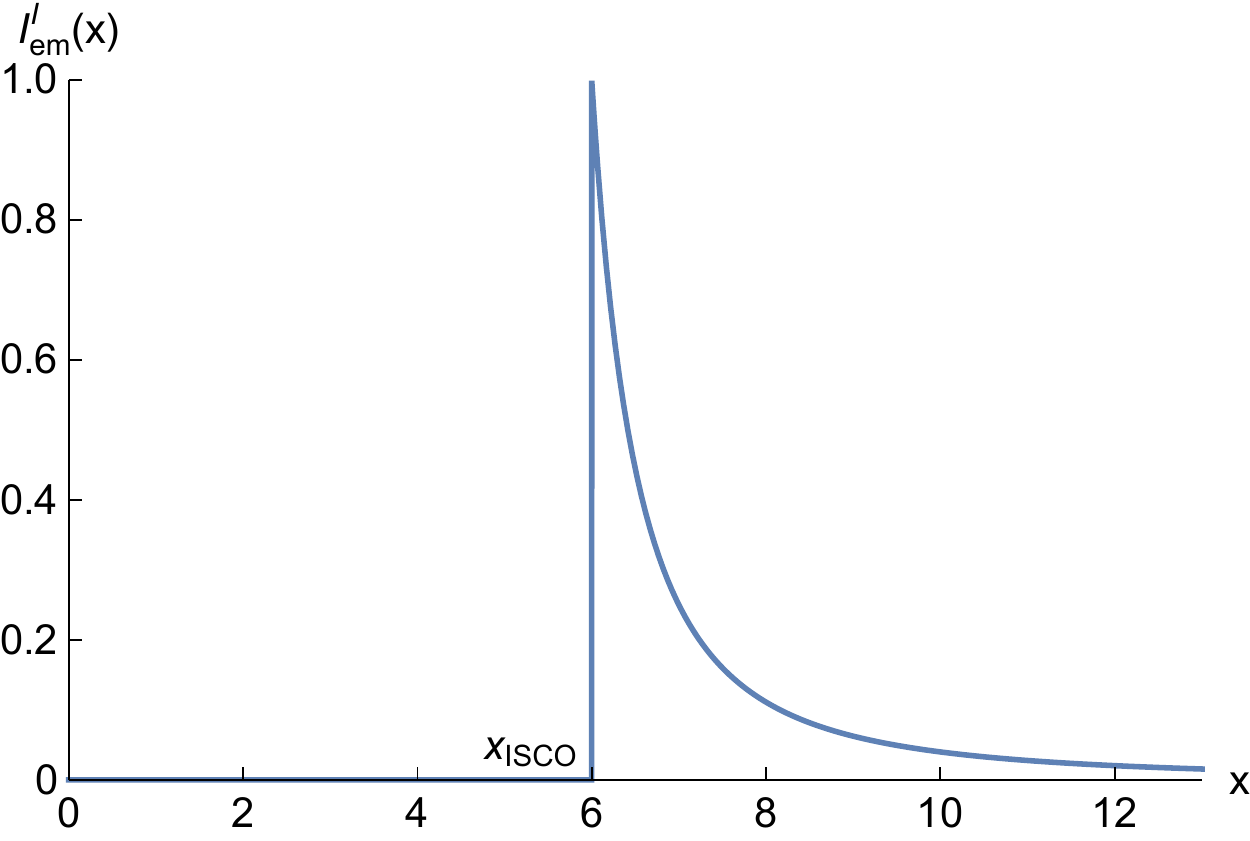}
\includegraphics[width=5.9cm,height=4.8cm]{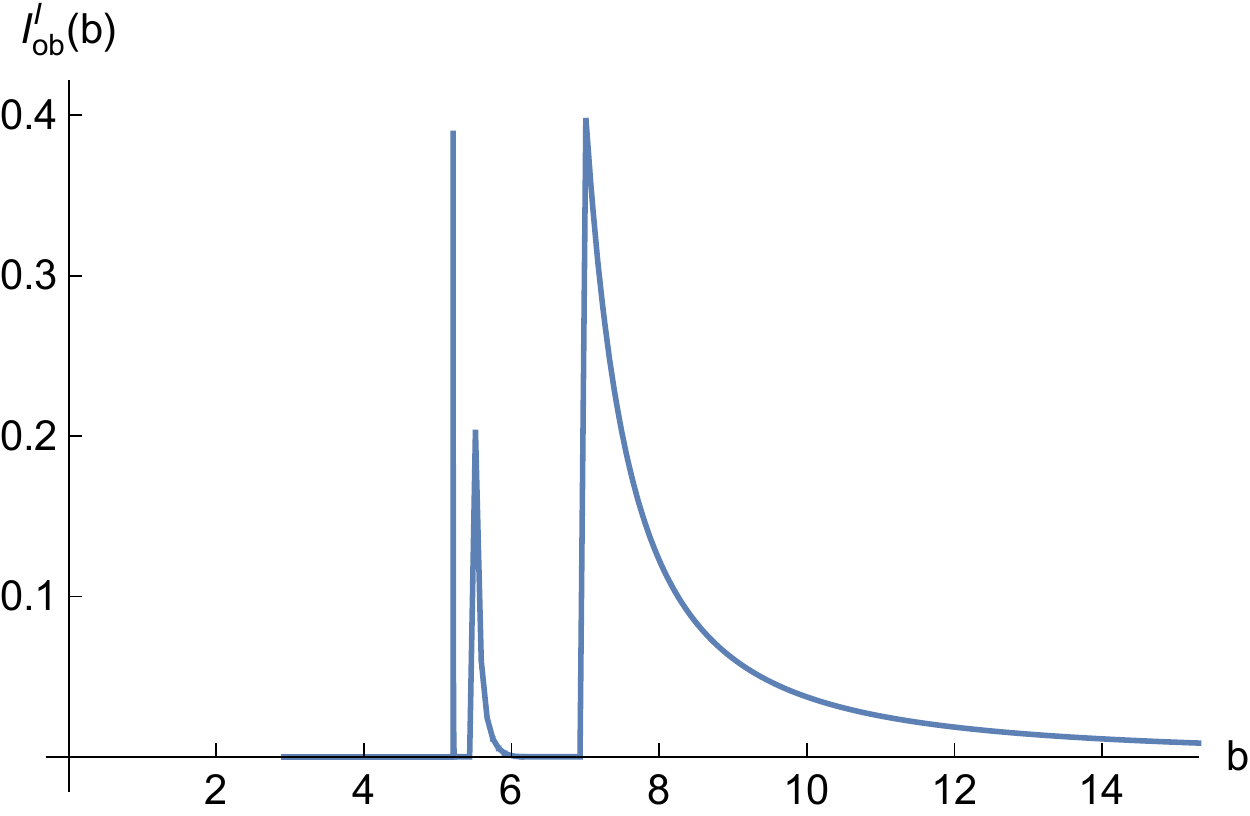}
\includegraphics[width=5.9cm,height=4.8cm]{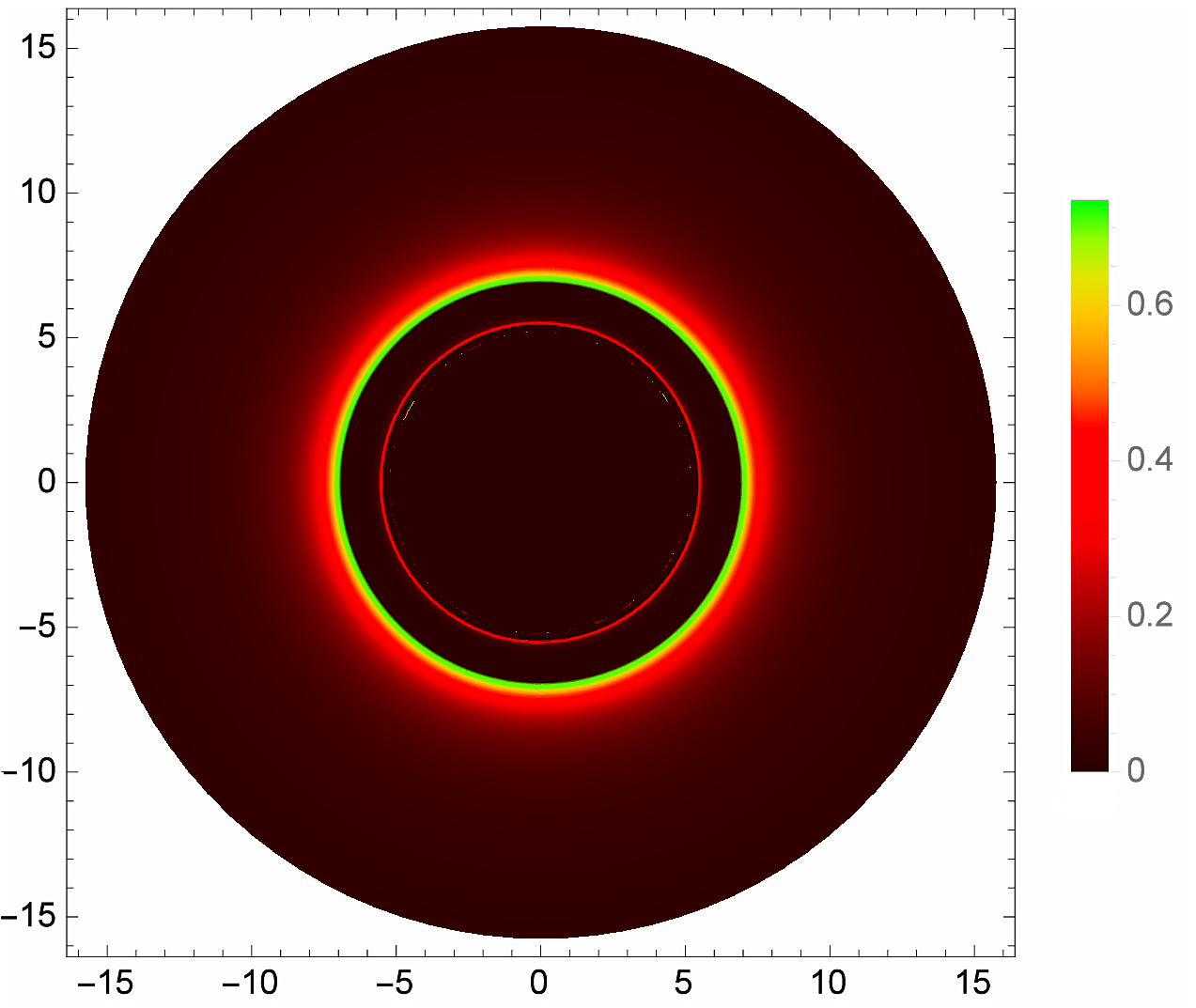}
\includegraphics[width=5.9cm,height=4.8cm]{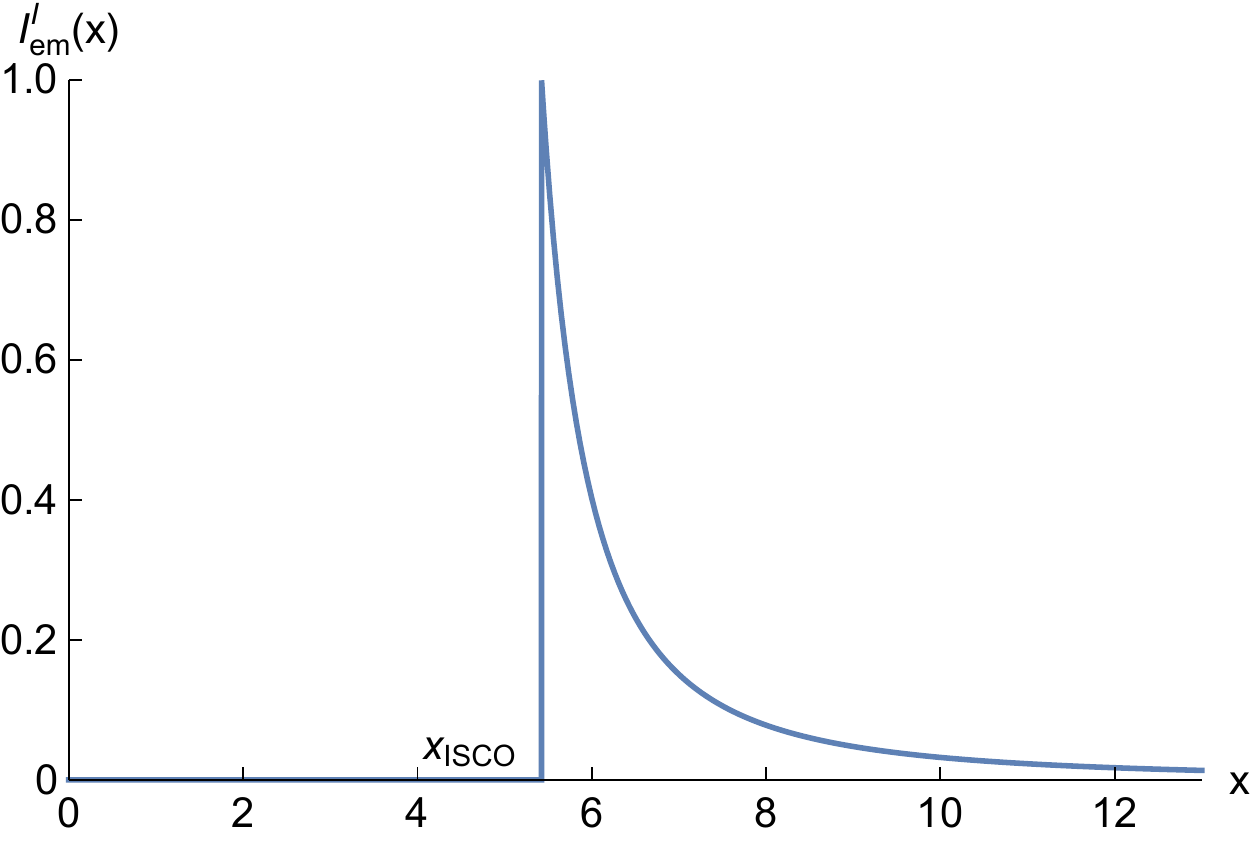}
\includegraphics[width=5.9cm,height=4.8cm]{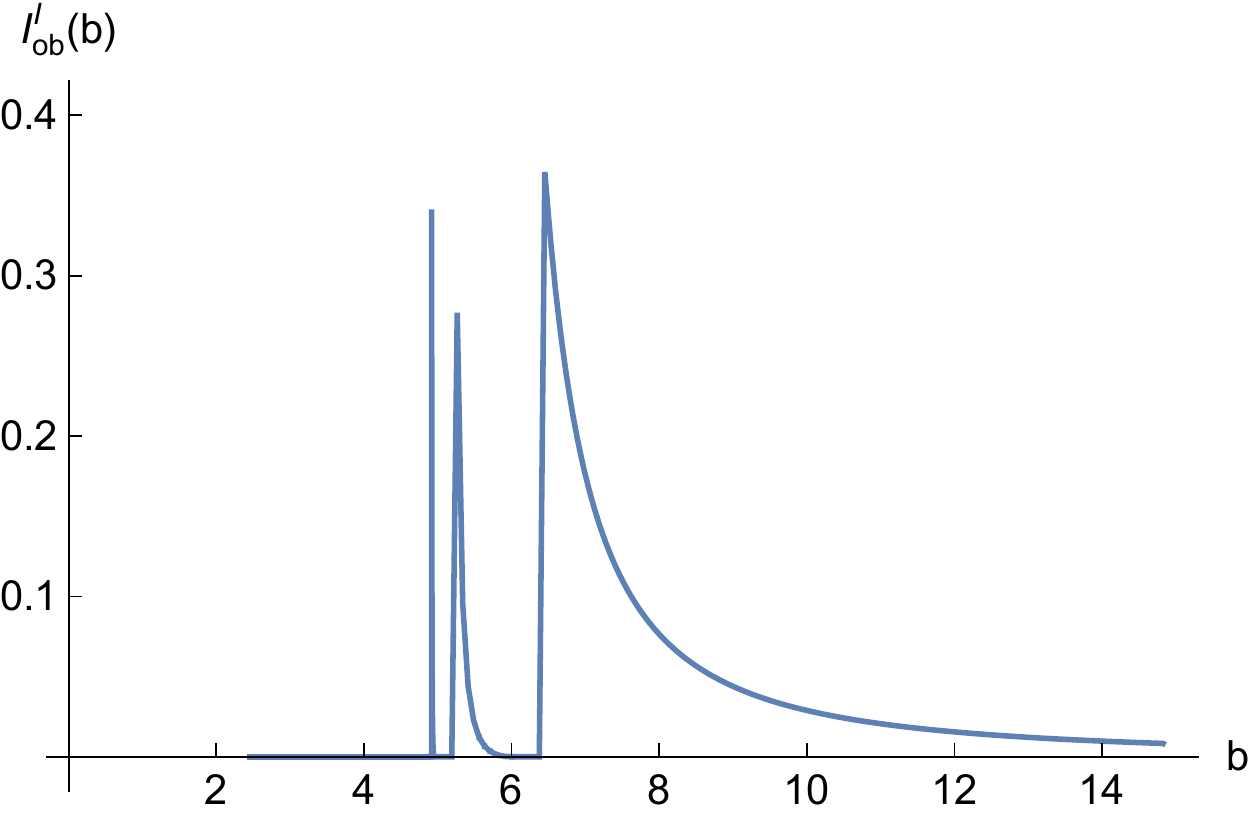}
\includegraphics[width=5.9cm,height=4.8cm]{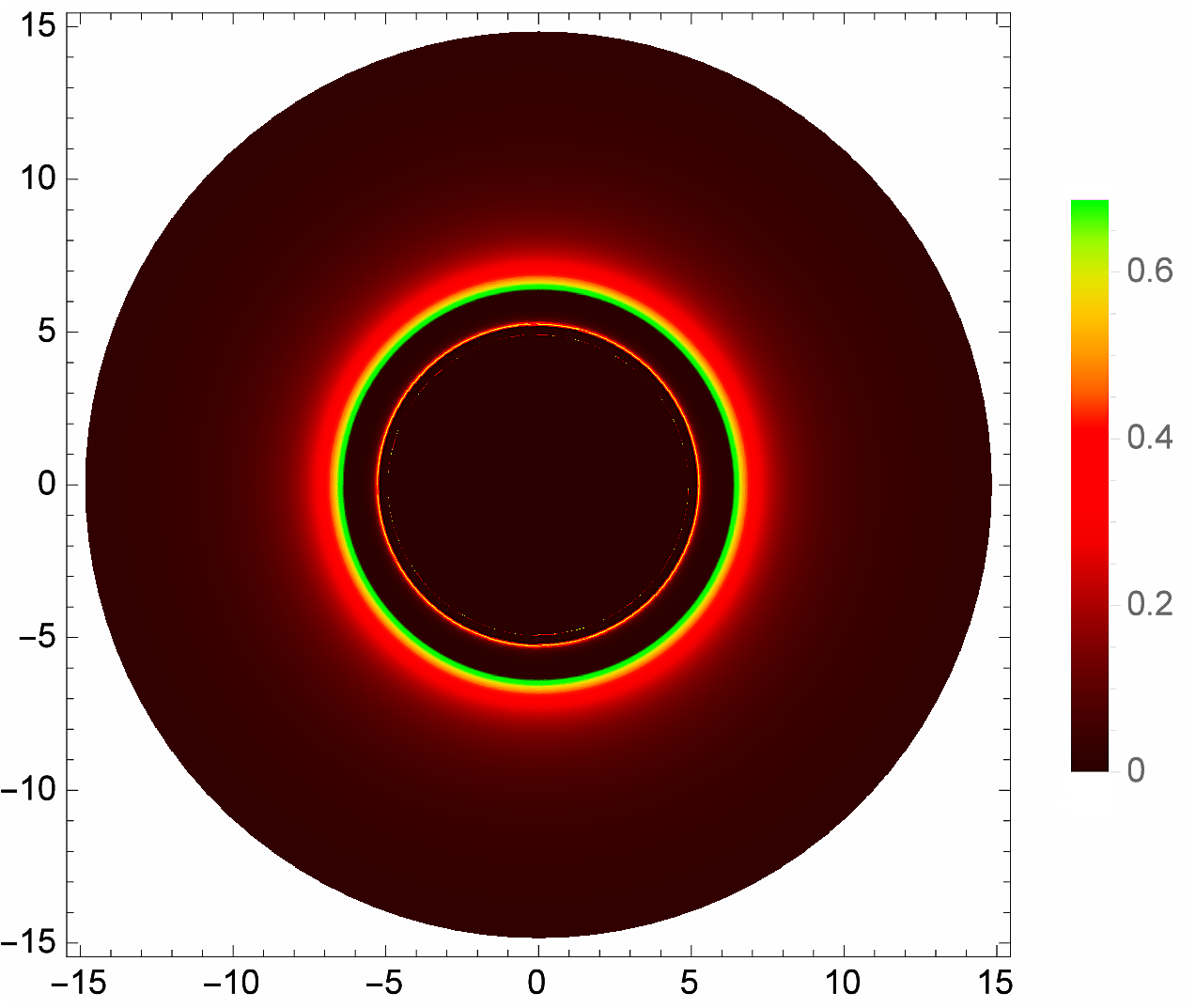}
\caption{The emitted luminosity (left), the observed one (middle) and the optical appearance (right) for the Schwarzschild black hole (top figures) and the two-horizons black hole with $a=2/3$ (bottom figures) for Model I. In this model $x_{isco}$ denotes the innermost stable circular orbit for time-like observers, which is located  (in units of $M=1$) at $x_{isco}=6$ for the Schwarzschild black hole and at $x_{isco} \approx 5.6$ for the two-horizons black hole. }
\label{fig:shadowBHcaseI}
\includegraphics[width=5.9cm,height=4.8cm]{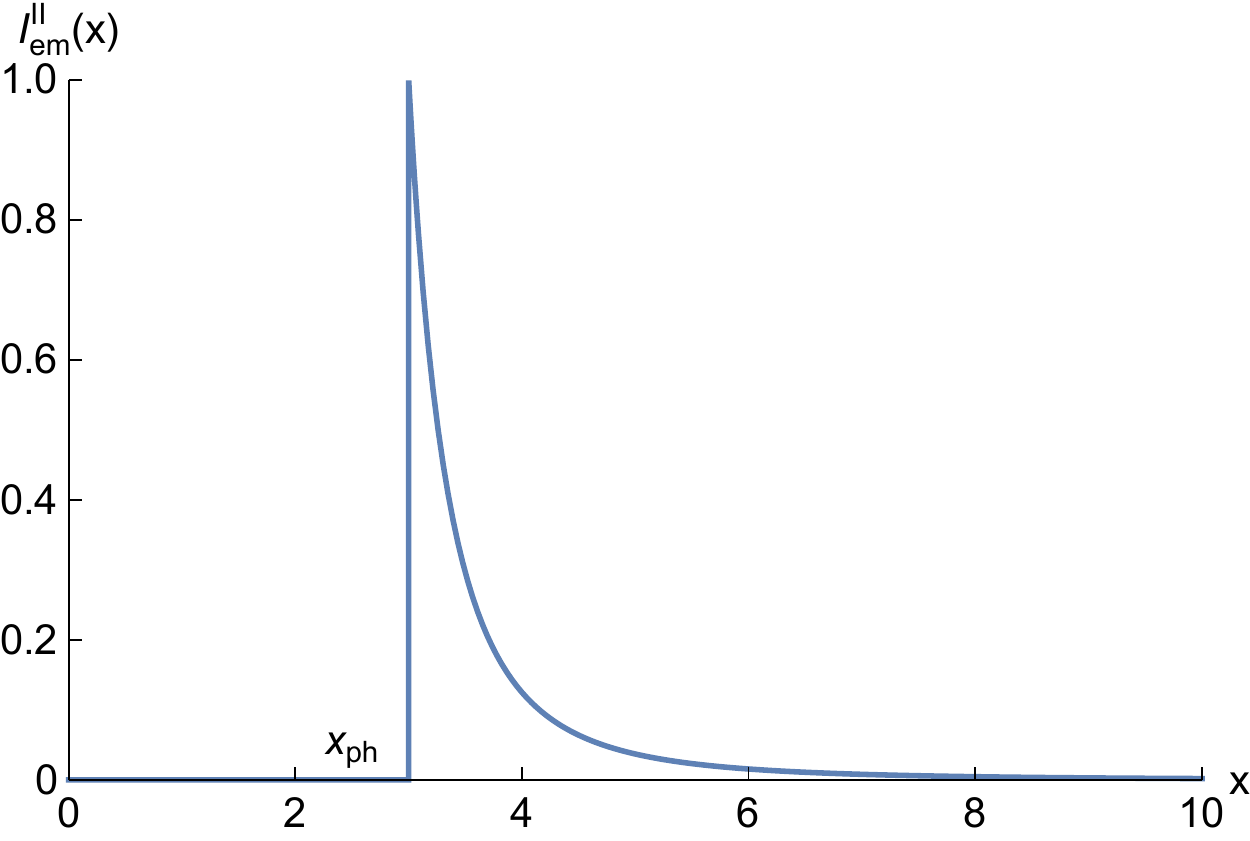}
\includegraphics[width=5.9cm,height=4.8cm]{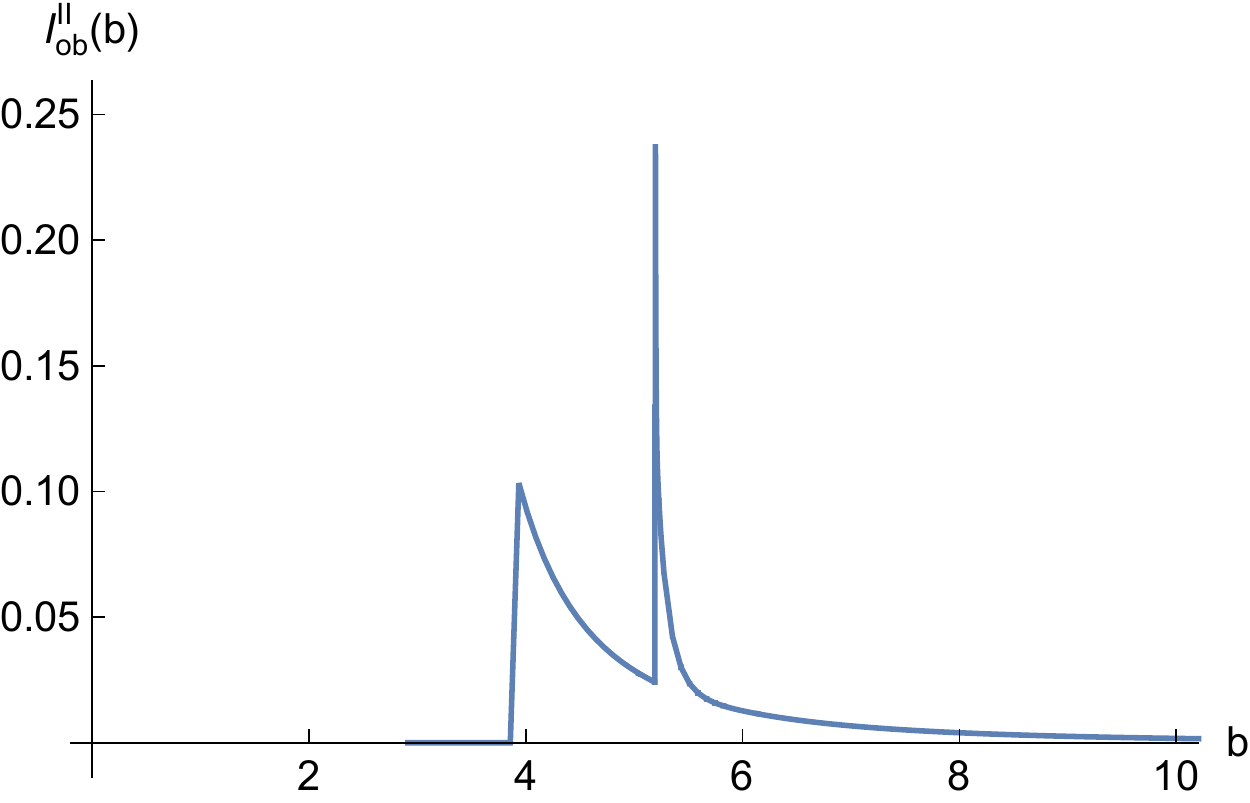}
\includegraphics[width=5.9cm,height=4.8cm]{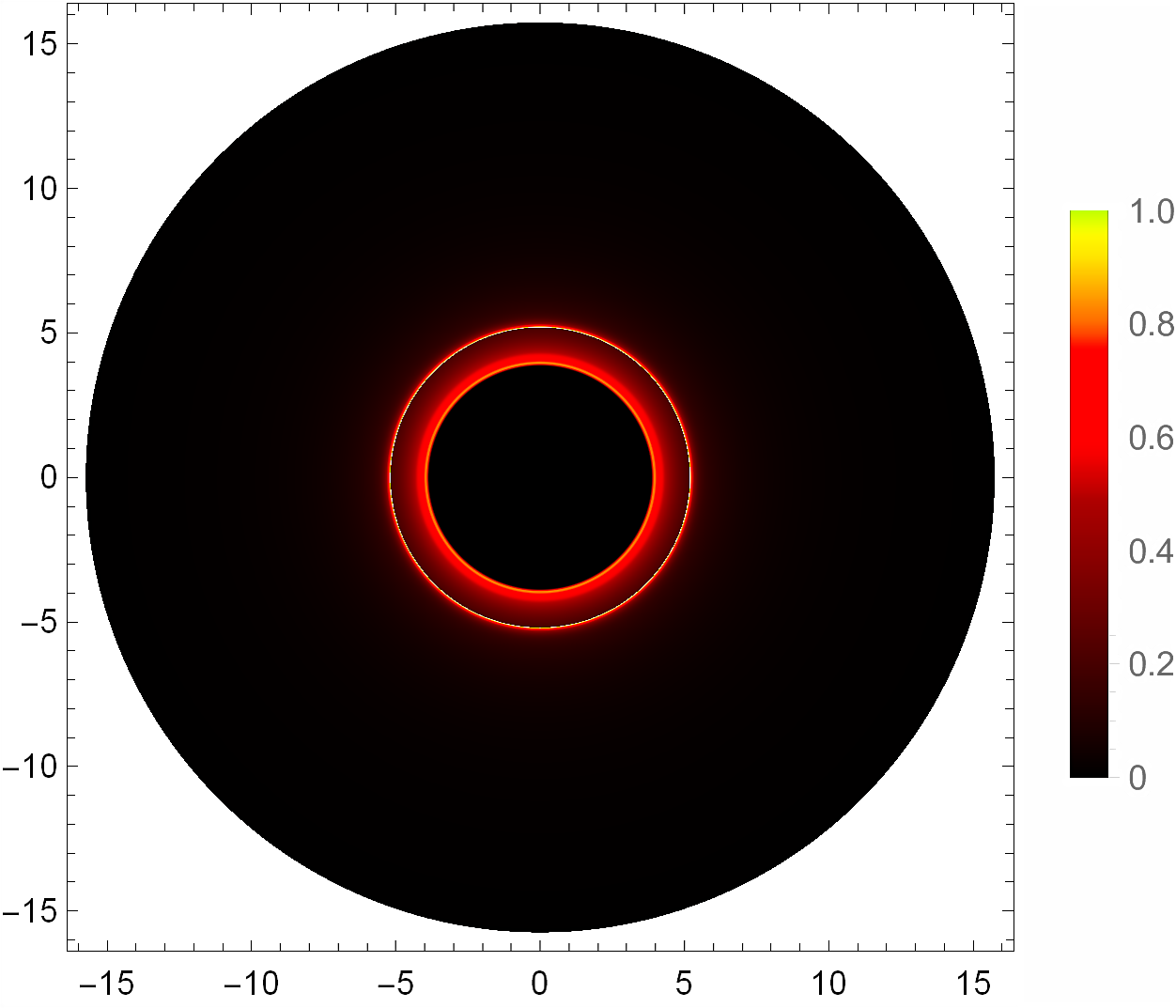}
\includegraphics[width=5.9cm,height=4.8cm]{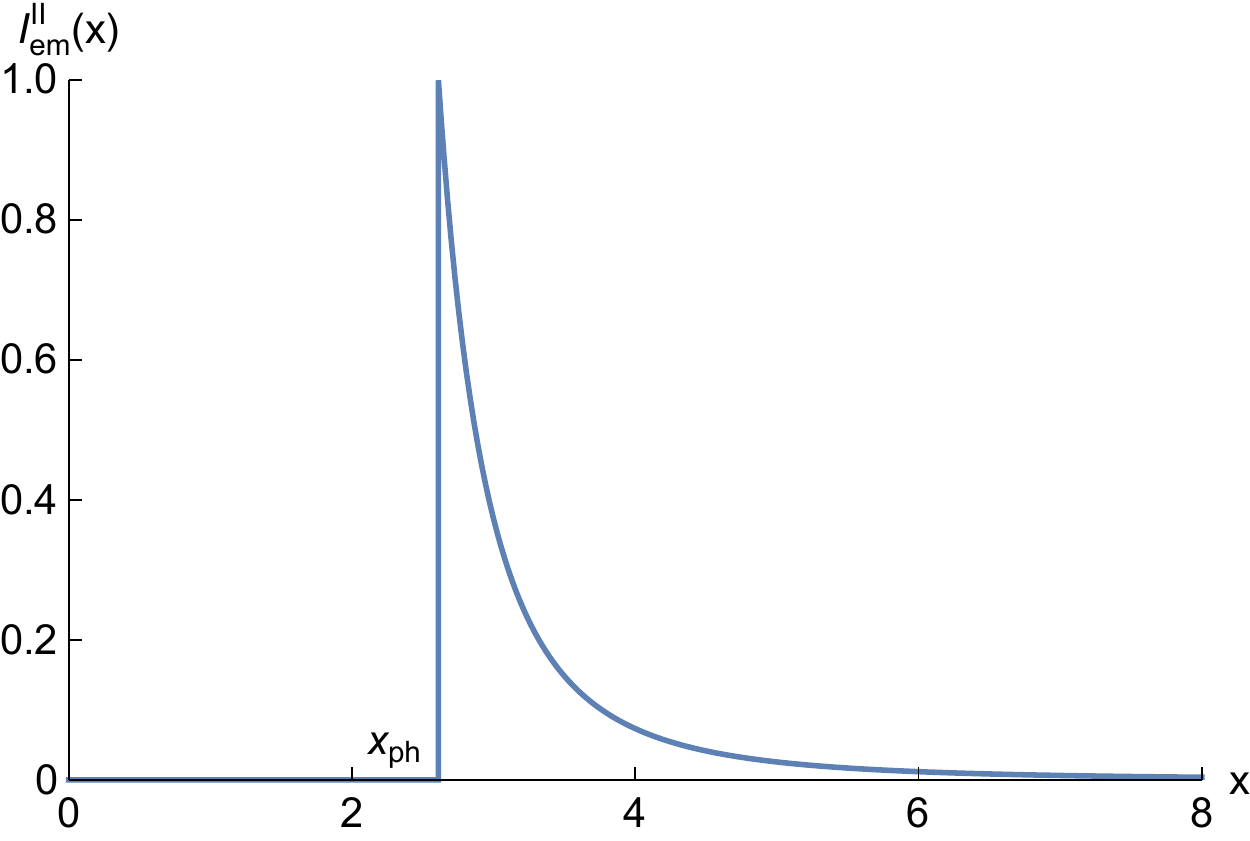}
\includegraphics[width=5.9cm,height=4.8cm]{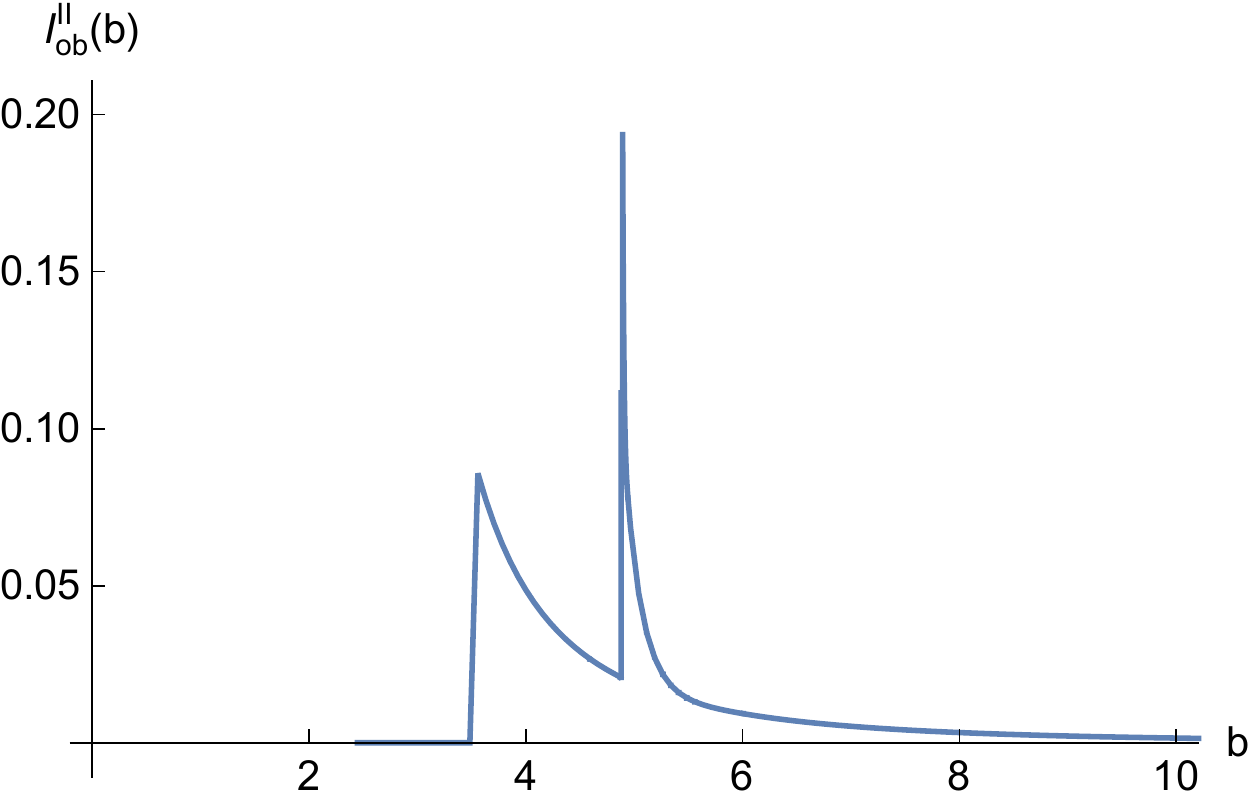}
\includegraphics[width=5.9cm,height=4.8cm]{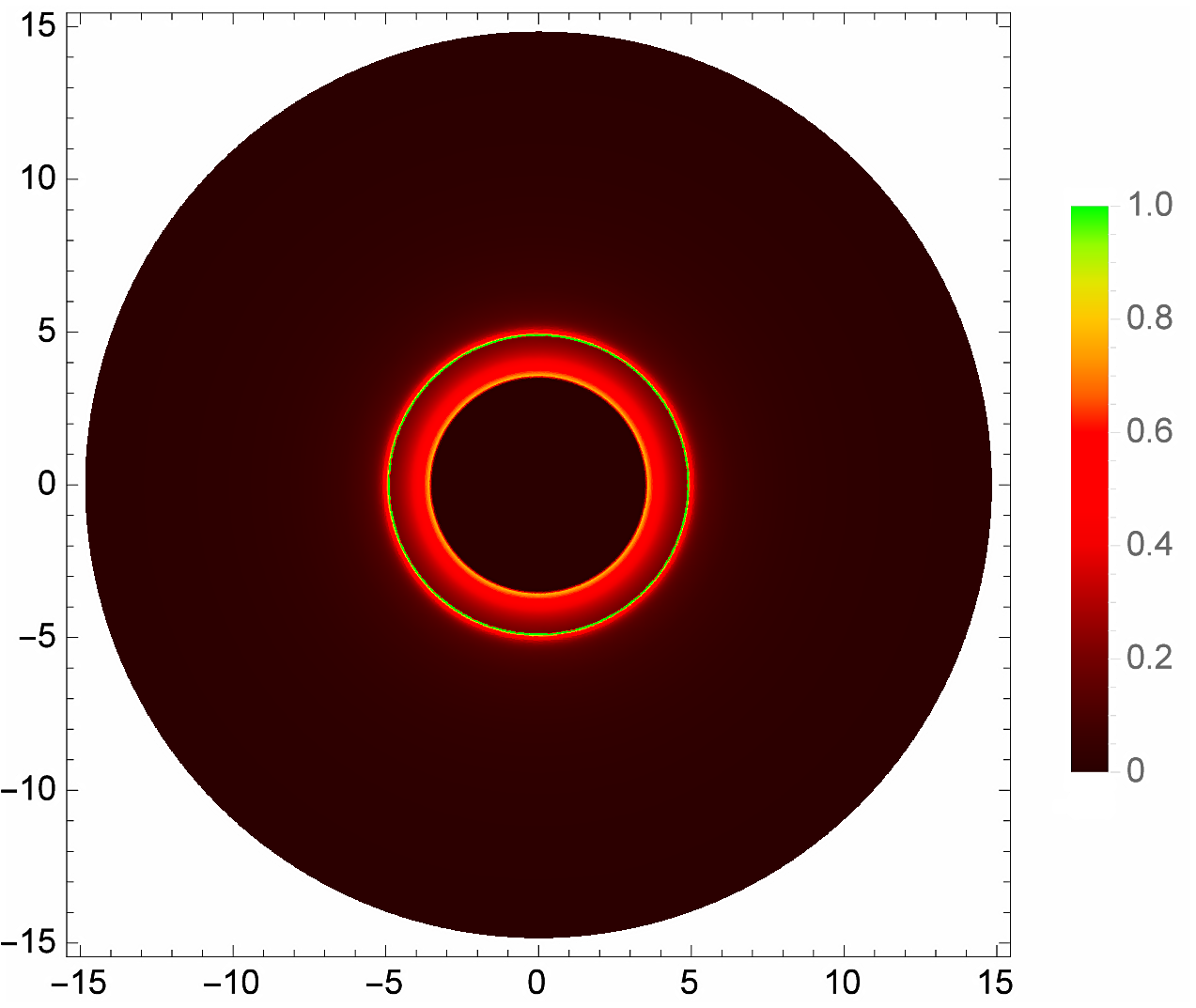}
\caption{The emitted luminosity (left), the observed one (middle) and the optical appearance (right) for the Schwarzschild black hole (top figures) and the two-horizons black hole with $a=2/3$ (bottom figures) for Model II. In this model $x_{ps}$ denotes the location of the photon sphere, which is located  (in units of $M=1$) at $x_{ps}=3$ for the Schwarzschild black hole and at $x_{ps} \approx 2.4$ for the two-horizons black hole.}
\label{fig:shadowBHcaseII}
\end{figure*}

\begin{figure*}[t!]
\includegraphics[width=5.9cm,height=4.8cm]{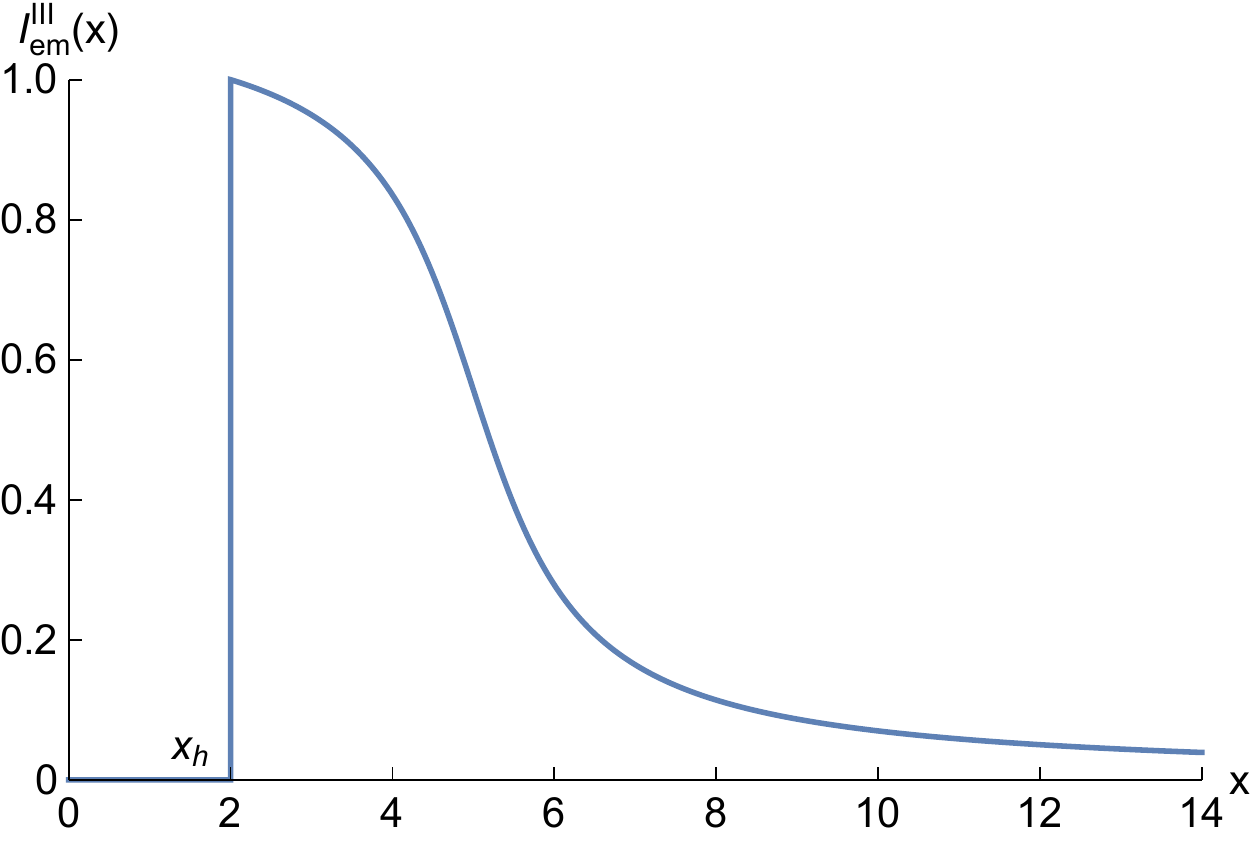}
\includegraphics[width=5.9cm,height=4.8cm]{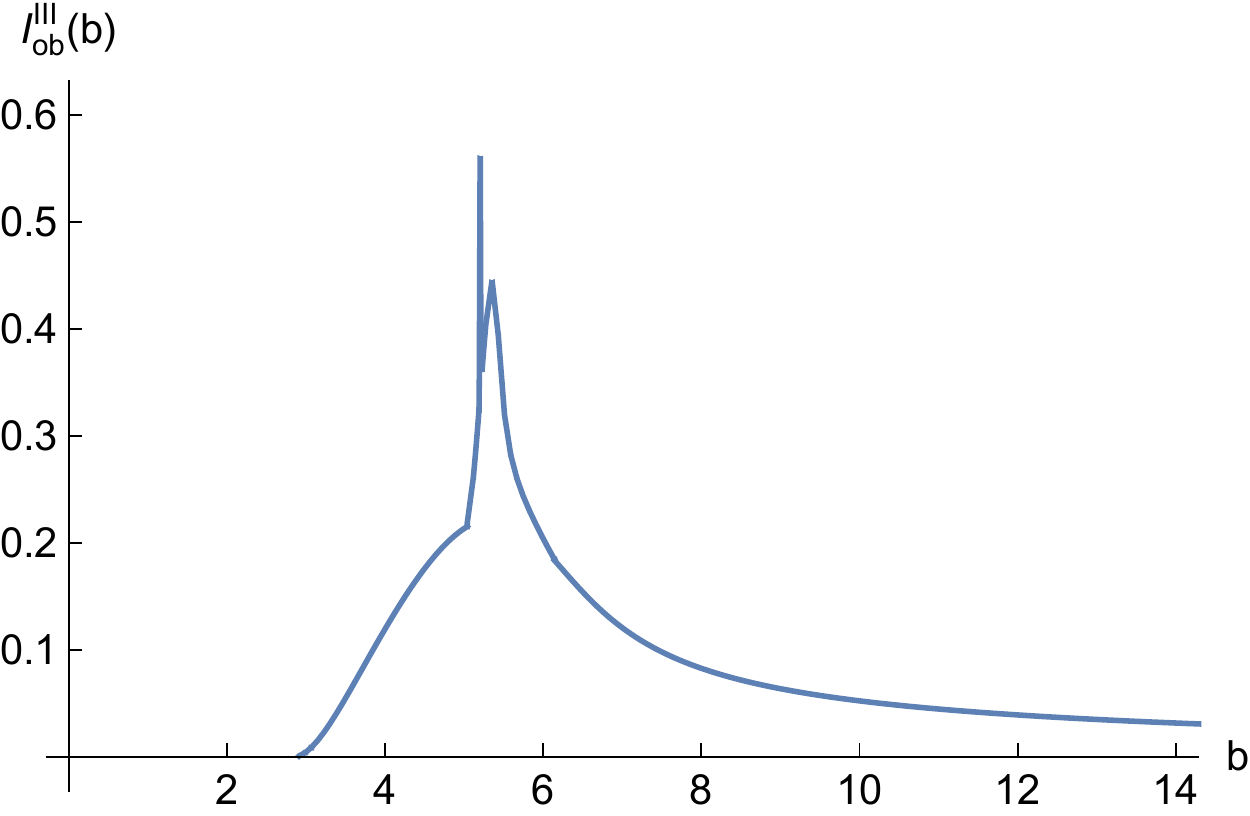}
\includegraphics[width=5.9cm,height=4.8cm]{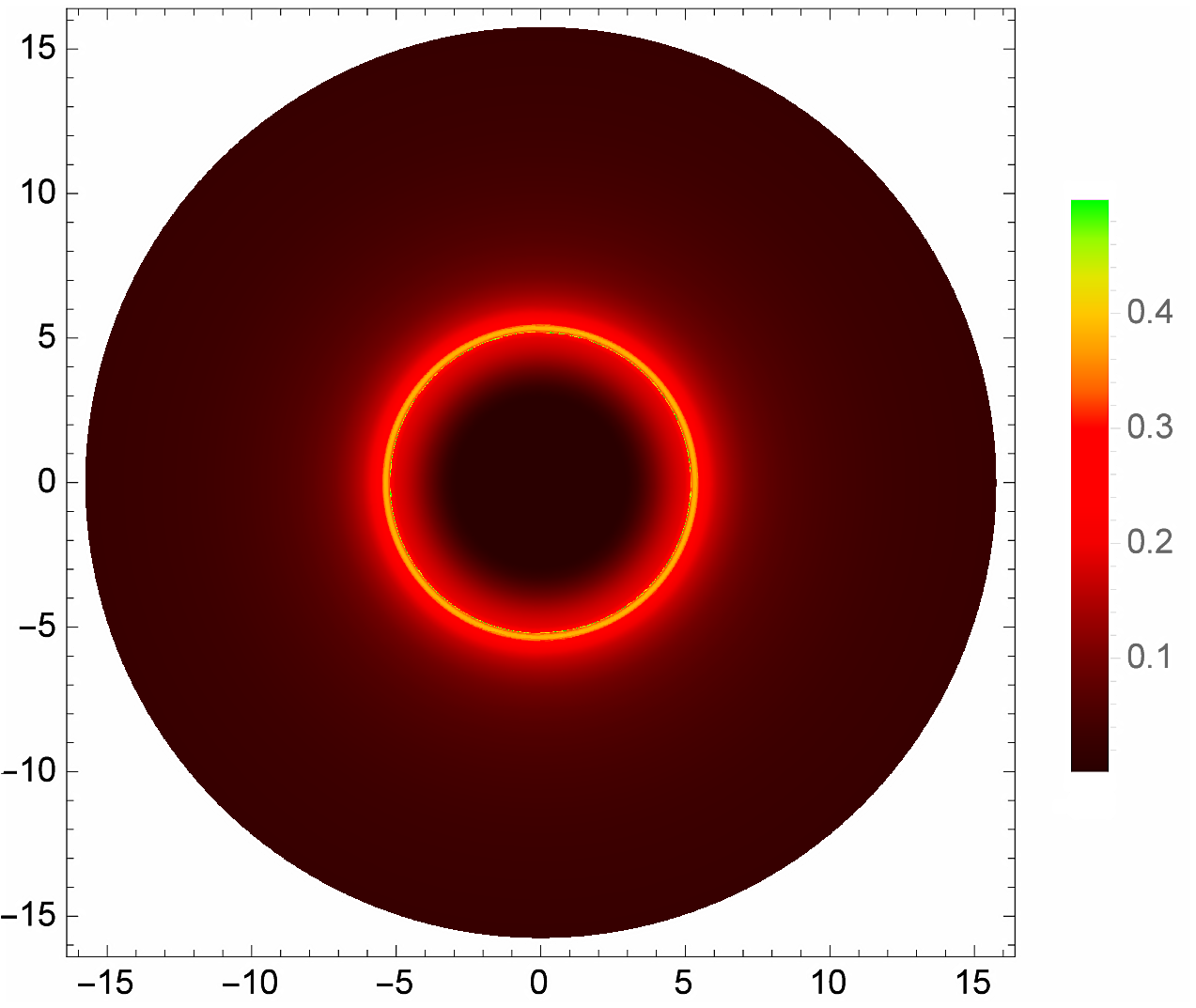}
\includegraphics[width=5.9cm,height=4.8cm]{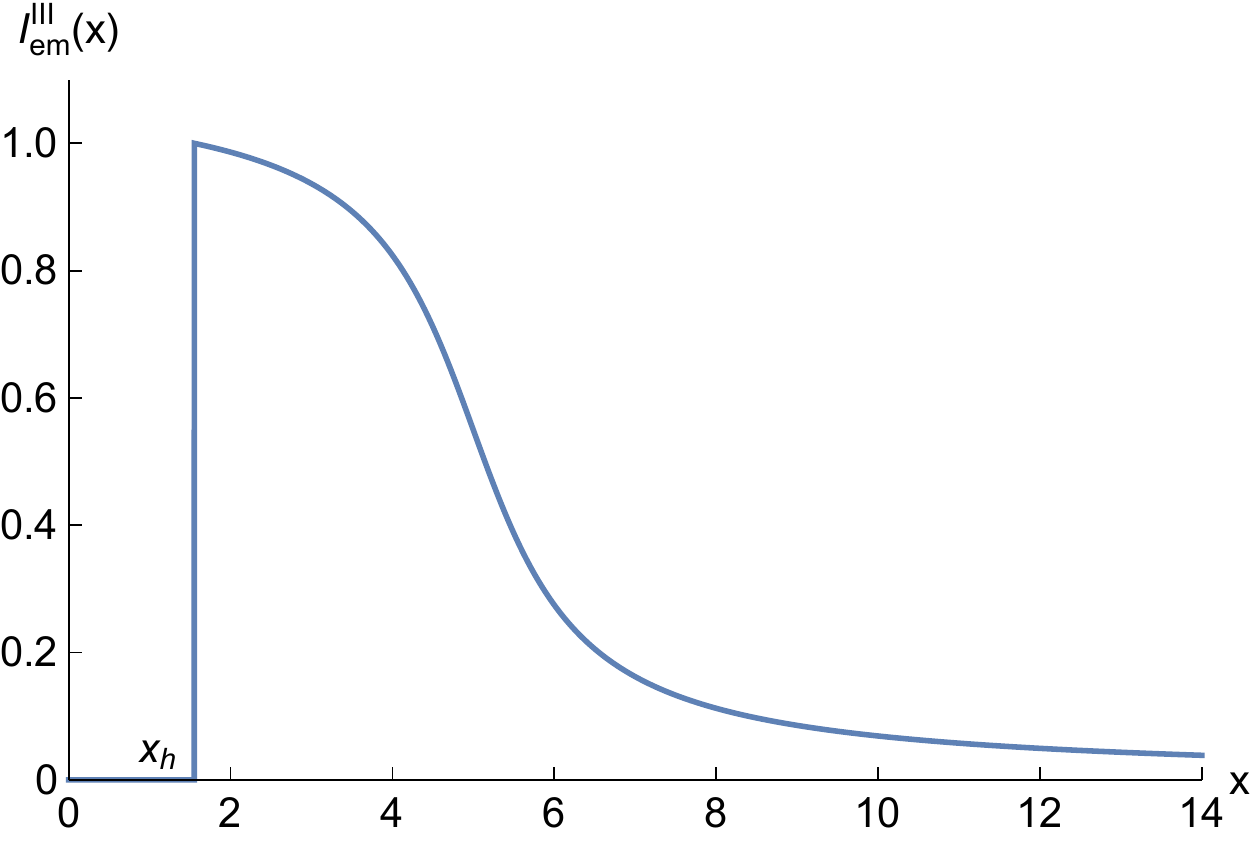}
\includegraphics[width=5.9cm,height=4.8cm]{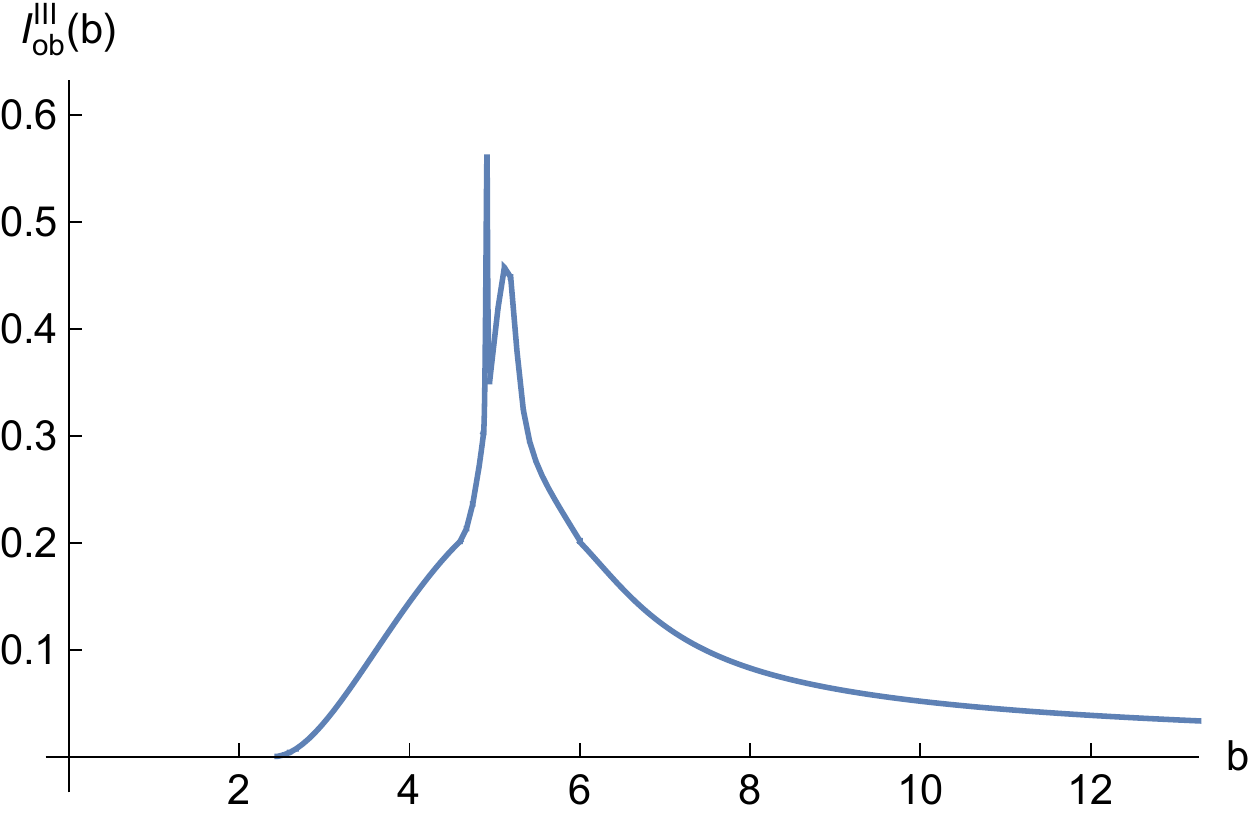}
\includegraphics[width=5.9cm,height=4.8cm]{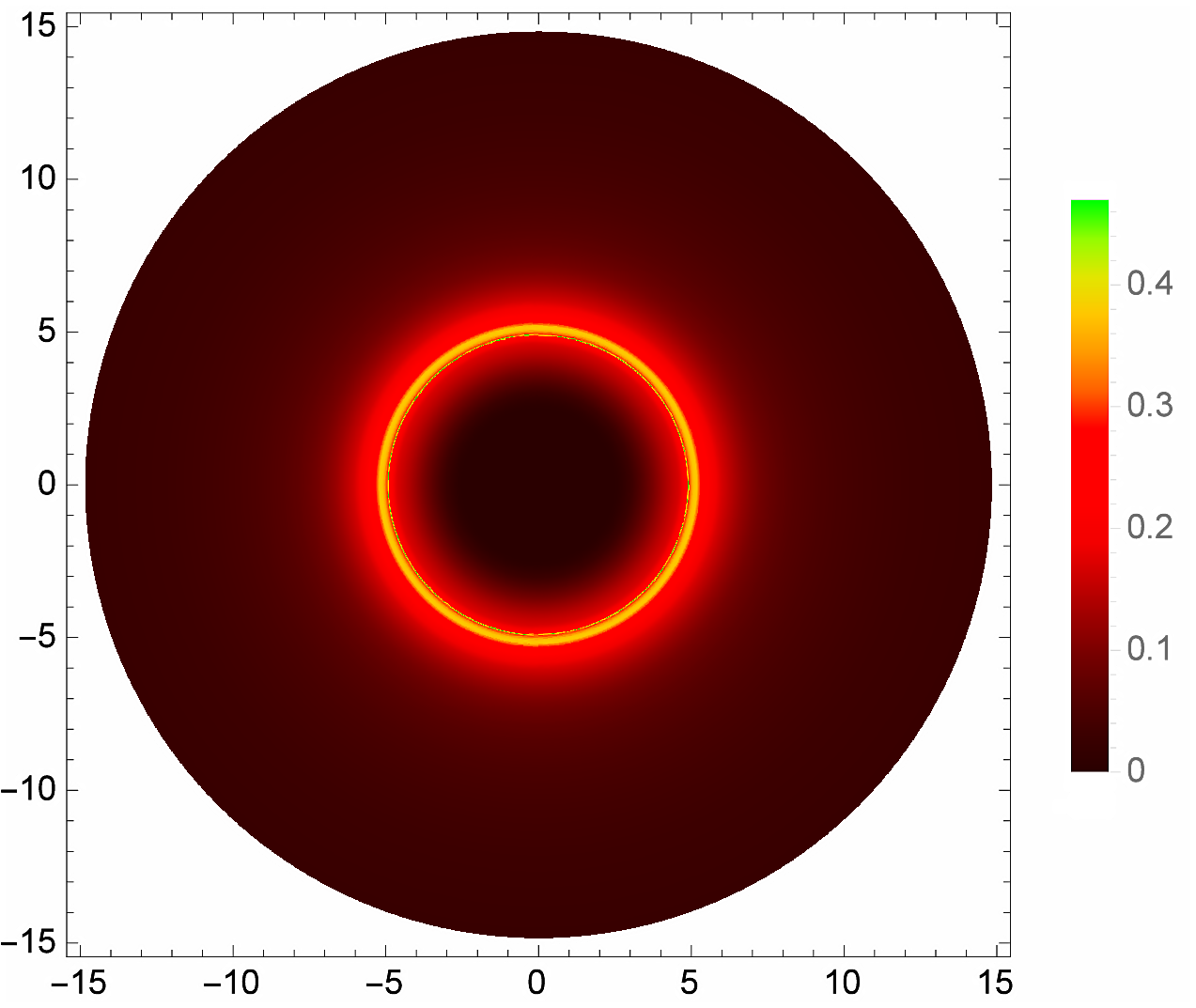}
\caption{The emitted luminosity (left), the observed one (middle) and the optical appearance (right) for the Schwarzschild black hole (top figures) and the two-horizons black hole with $a=2/3$ (bottom figures) for Model III. In this model $x_{h}$ denotes the location of the horizon, which is located (in units of $M=1$) at $x_{h}=2$ for the Schwarzschild solution and at $x_{h} \approx 1.55$ for the two-horizons black hole.}
\label{fig:shadowBHcaseIII}
\end{figure*}

To heat up, let us consider first the two-horizons black hole case, where only the outer photon sphere is accessible above the event horizon.

We first consider Model I, and depict in Fig. \ref{fig:shadowBHcaseI}  the emitted (left) and observed (middle) intensity profiles as well as the optical appearances (right) for the two-horizons black hole configurations with $a=2/3$ (bottom) as compared to the one of the Schwarzschild solution (top). In this Model I, the fact that the emission starts at the innermost stable circular orbit for time-like observers allows the (gravitationally redshifted) observed luminosity to clearly isolate the impact parameter regions associated to the direct, lensed and photon ring emissions. This is translated into a clean view of the three kinds of light rings in the optical appearances image, with the direct emission largely dominating the total luminosity under a bright extended lump of radiation enclosing a thinner and dimmer ring (the lensed emission) and ending in an even thinner photon ring which is barely visible at naked eye. The modifications introduced by the two-horizons black hole as compared to the Schwarzschild solution moderately increase the width and luminosity of the lensed and photon ring emissions, thanks to the enhanced impact parameter region discussed in Sec. \ref{sec:Ray}.

In Model II, which is depicted in Fig. \ref{fig:shadowBHcaseII}, the fact that the inner edge of the accretion disk extends down to the location of the critical curve itself, enables the direct emission via the gravitational redshift correction to pierce well inside the critical impact factor region and become the dominant contribution there, while for larger impact parameter values the combined lensed and photon ring emissions occurring roughly at the same location produce a large spike in the observed emission, superimposed with the direct emission there. The net result in the optical appearance of this object within this Model II is a wide region of luminosity enclosing a thinner bright ring which is visible at the external part of it, an effect which is significantly enlarged in the two-horizons black hole case as compared to the Schwarzschild one.

Finally, in Model III, depicted in Fig. \ref{fig:shadowBHcaseIII}, since the inner edge of the disk extends all the way down to the event horizon this translates into a much wider region of luminosity in the observed emission, thanks to the stretching of the direct emission to a larger distance. The photon ring and lensed emissions appear now as two separated but superimposed spikes with the direct emission. As a consequence of these features, the optical appearance shows a much wider region of luminosity fuelled by the direct emission, enclosing a wide ring right on the middle of it, and another (dimmer) one right on the inner boundary of the latter. We point out that this effect of the superimposed rings is much  brighter and noticeable in the two-horizons black hole case than in the Schwarzschild solution.

\subsection{The traversable wormhole having two photon spheres}

\begin{figure*}[t!]
\includegraphics[width=5.9cm,height=5.2cm]{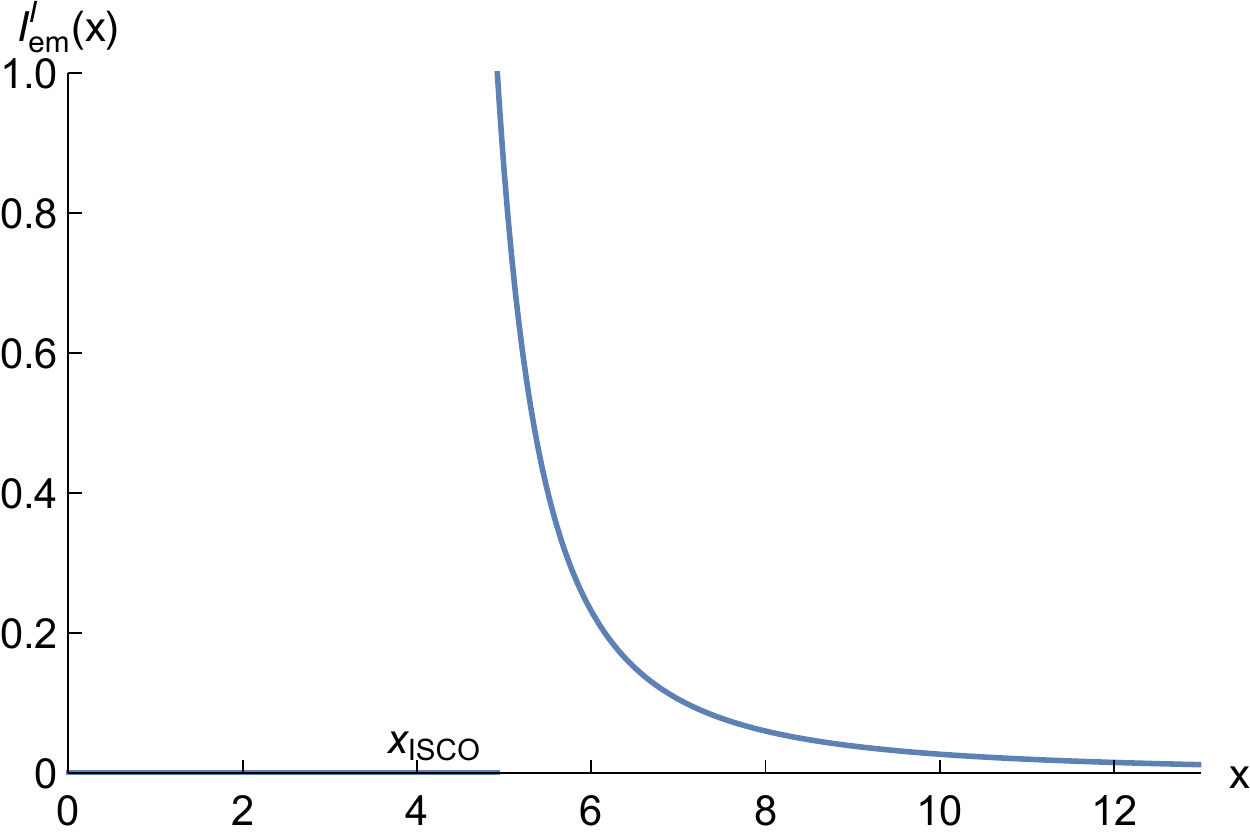}
\includegraphics[width=5.9cm,height=5.2cm]{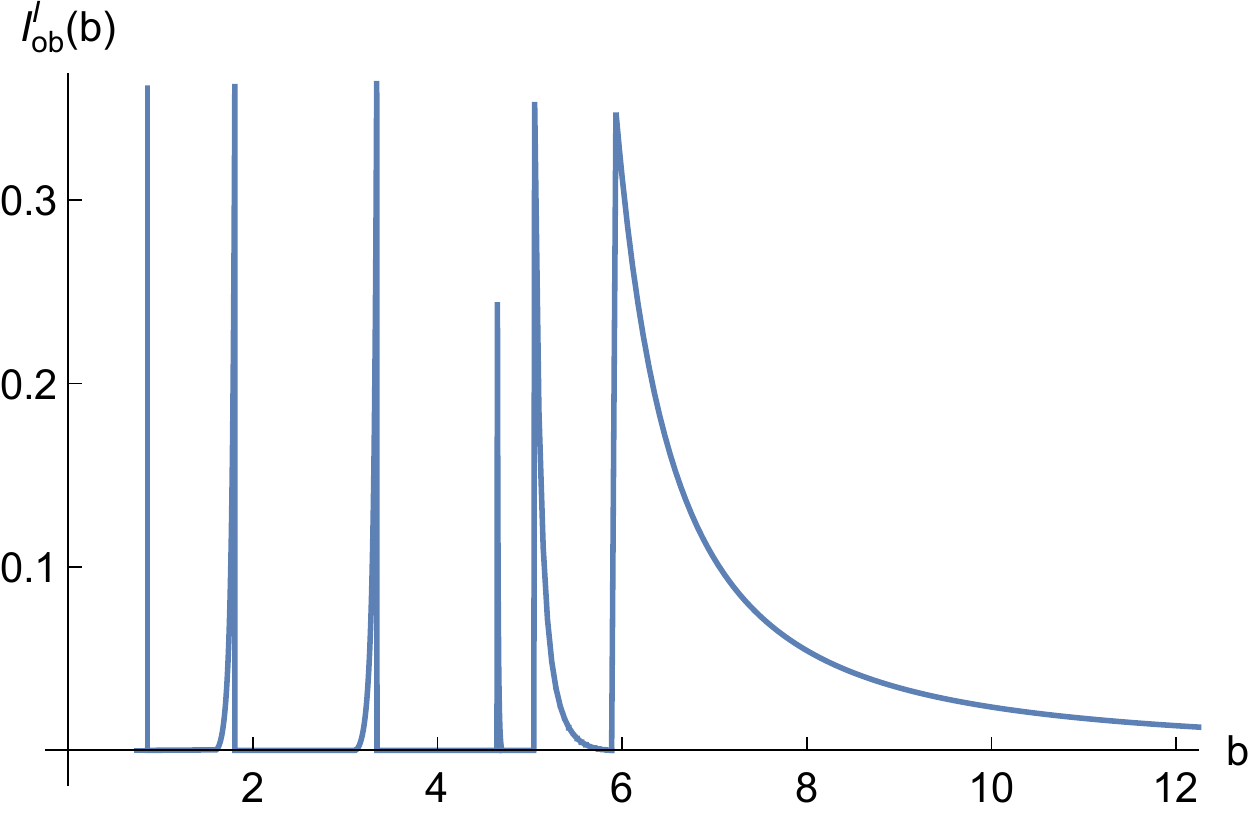}
\includegraphics[width=5.9cm,height=5.2cm]{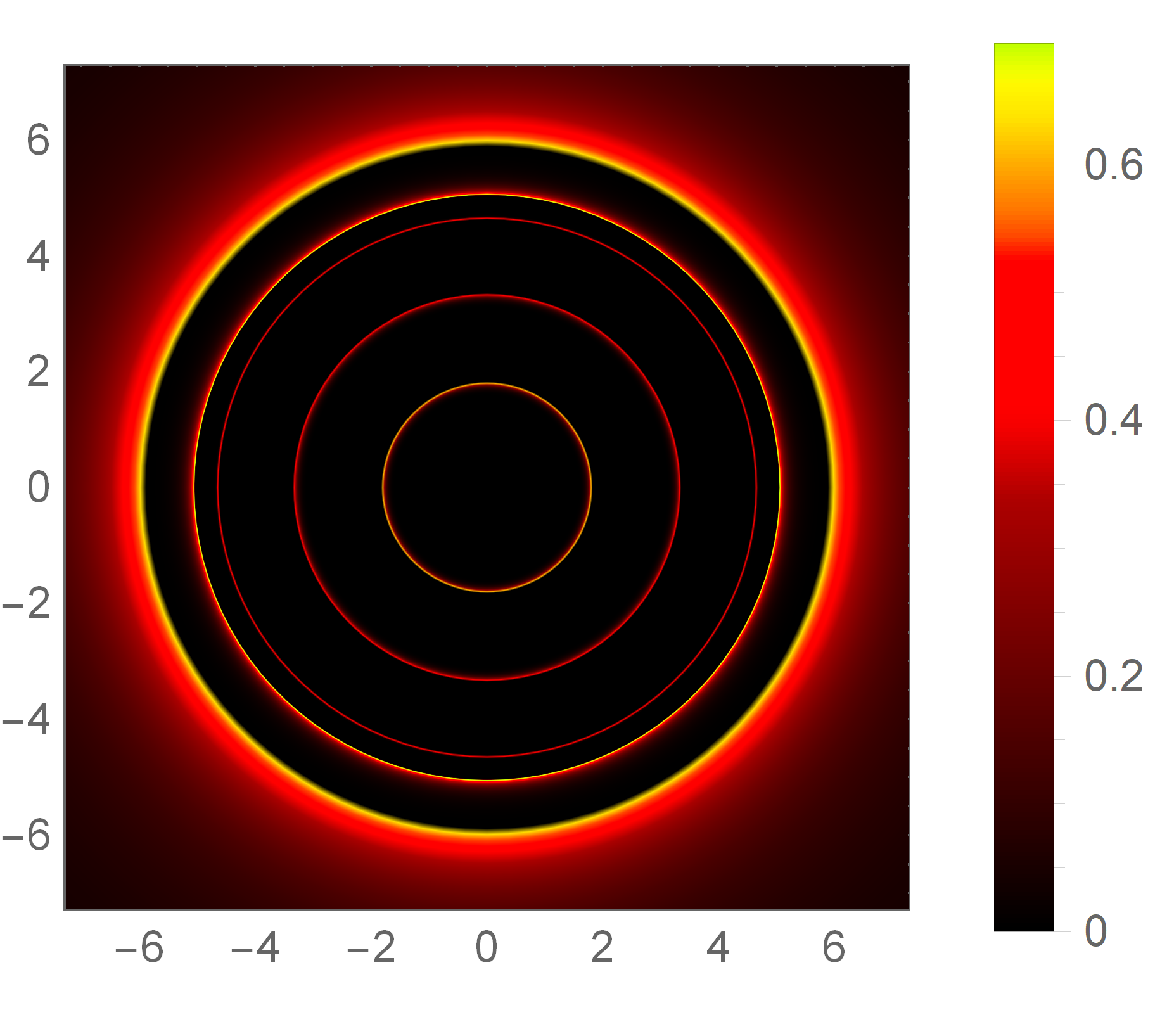}
\includegraphics[width=5.9cm,height=5.2cm]{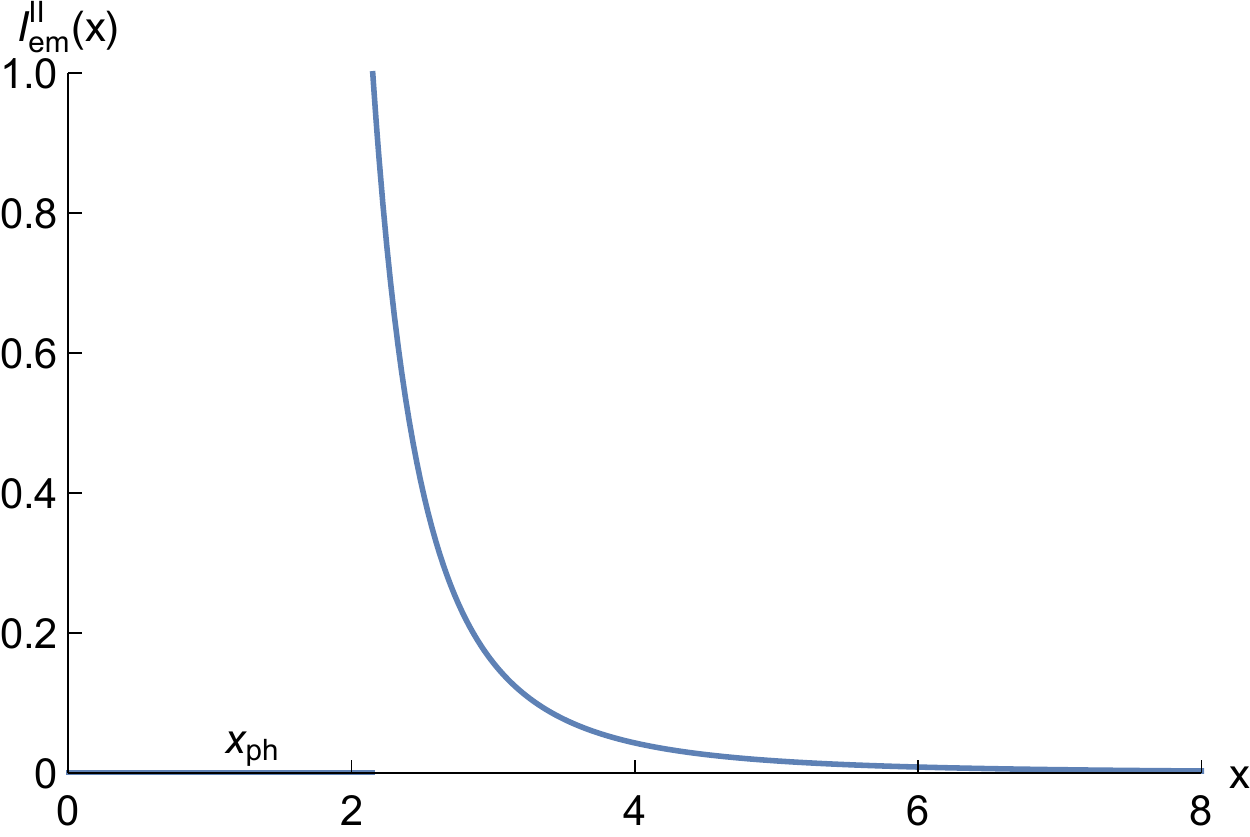}
\includegraphics[width=5.9cm,height=5.2cm]{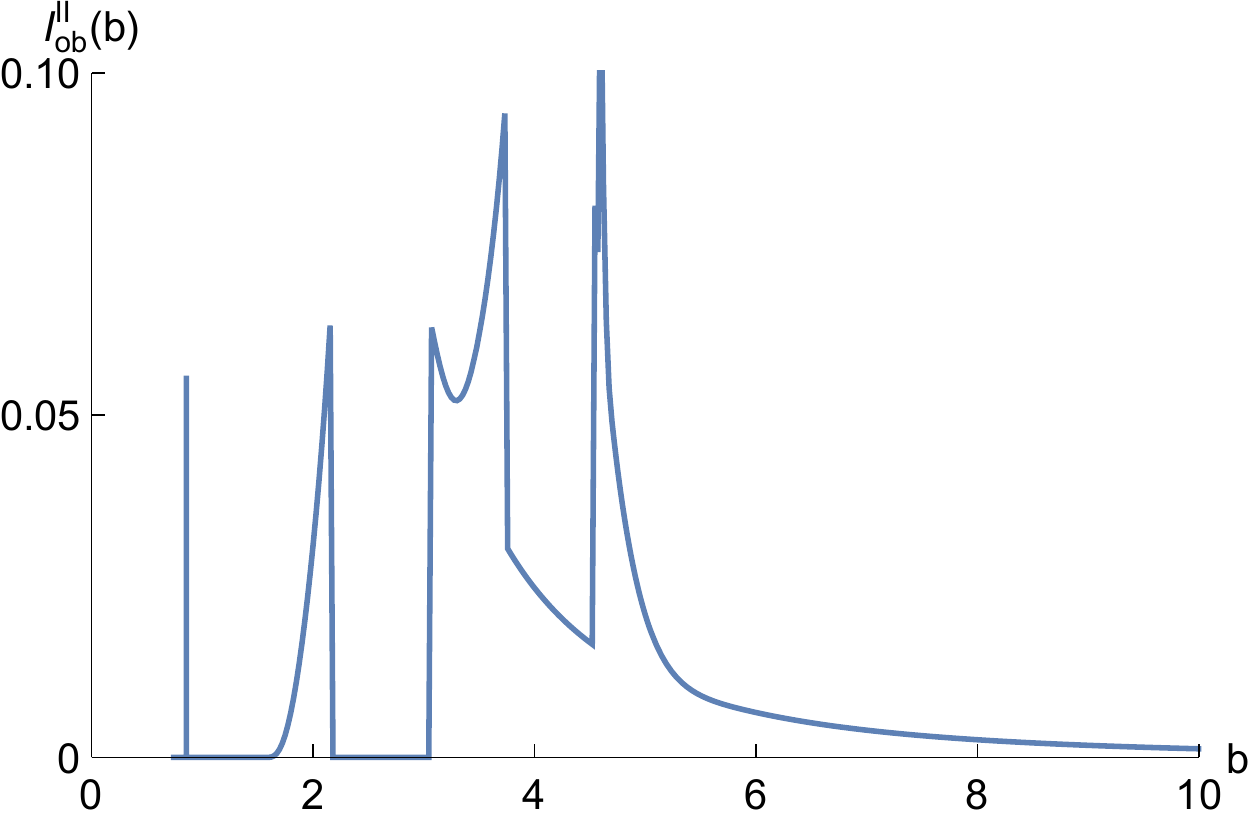}
\includegraphics[width=5.9cm,height=5.2cm]{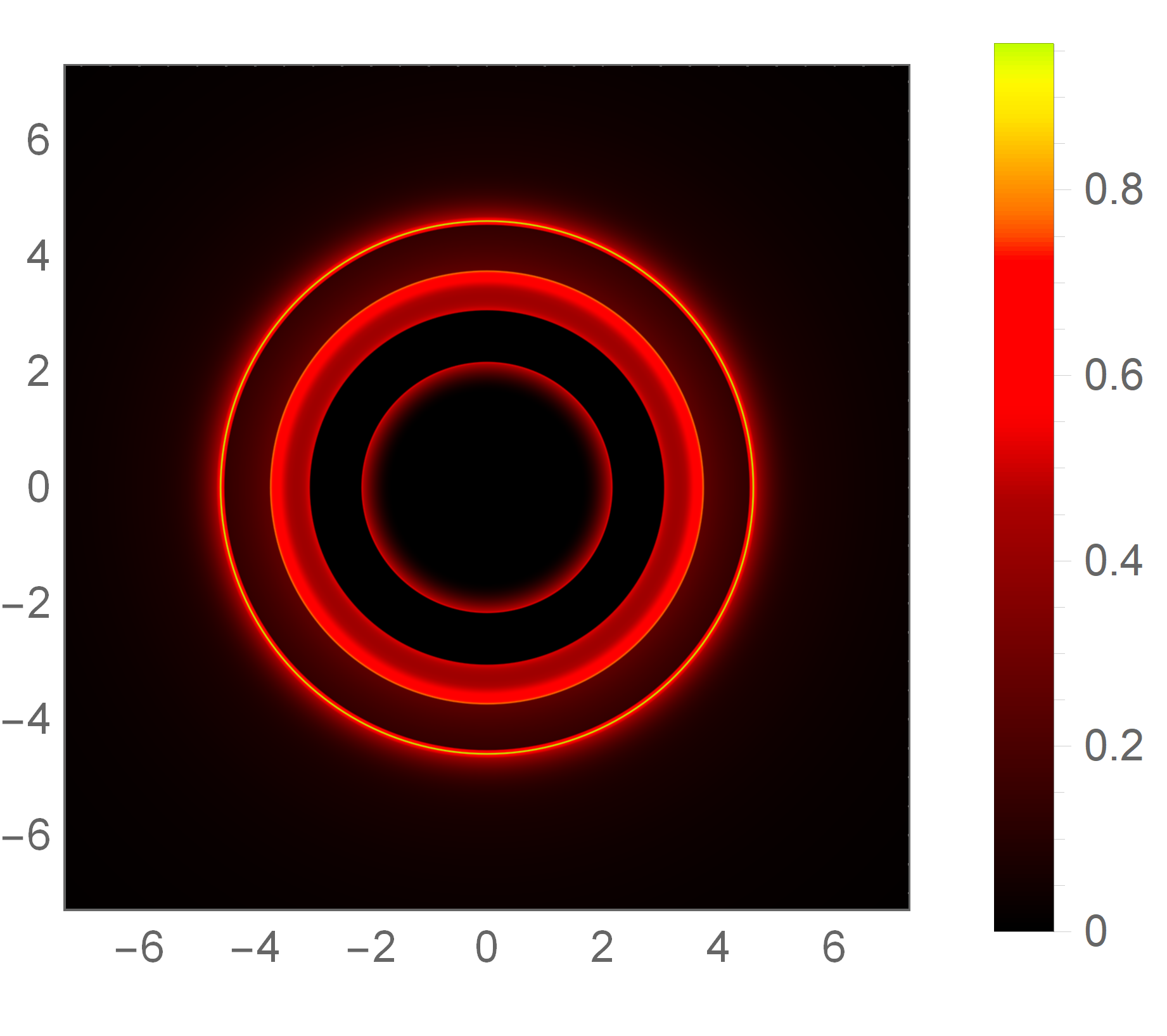}
\includegraphics[width=5.9cm,height=5.2cm]{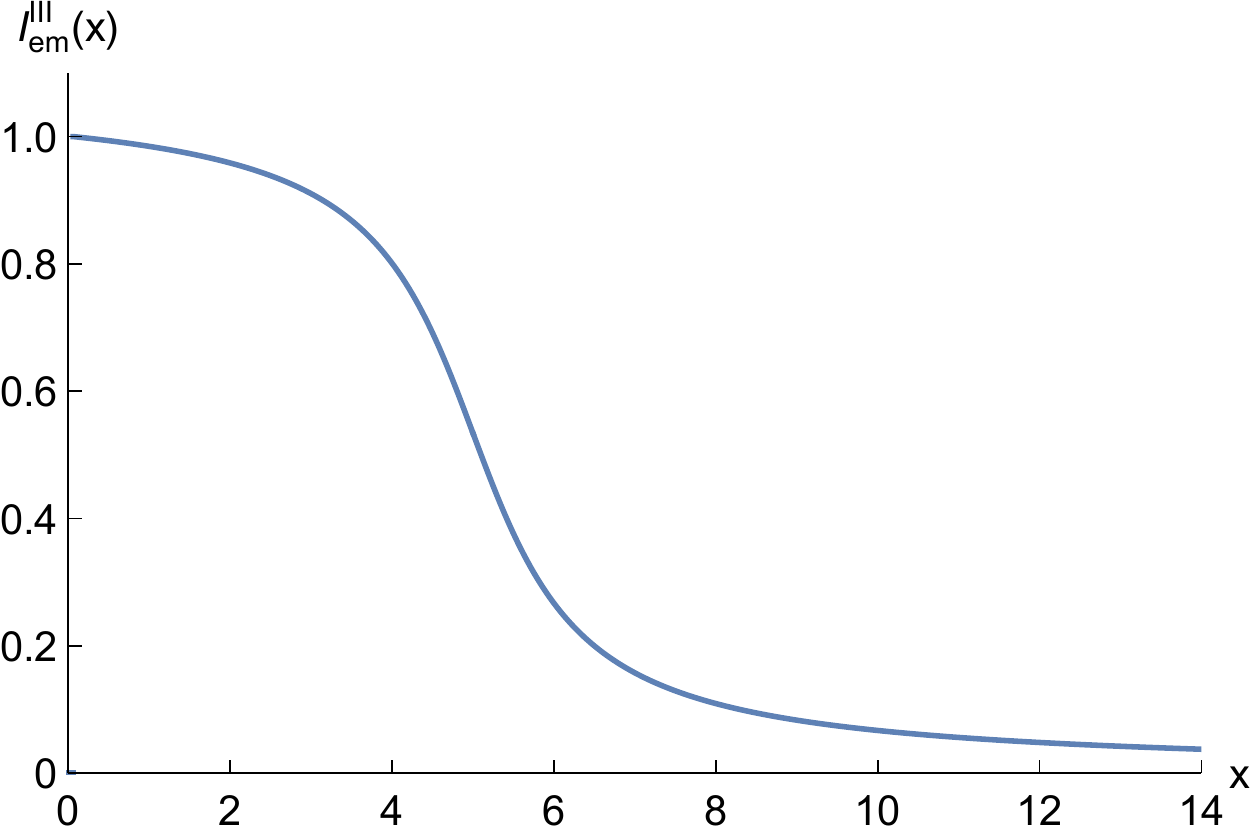}
\includegraphics[width=5.9cm,height=5.2cm]{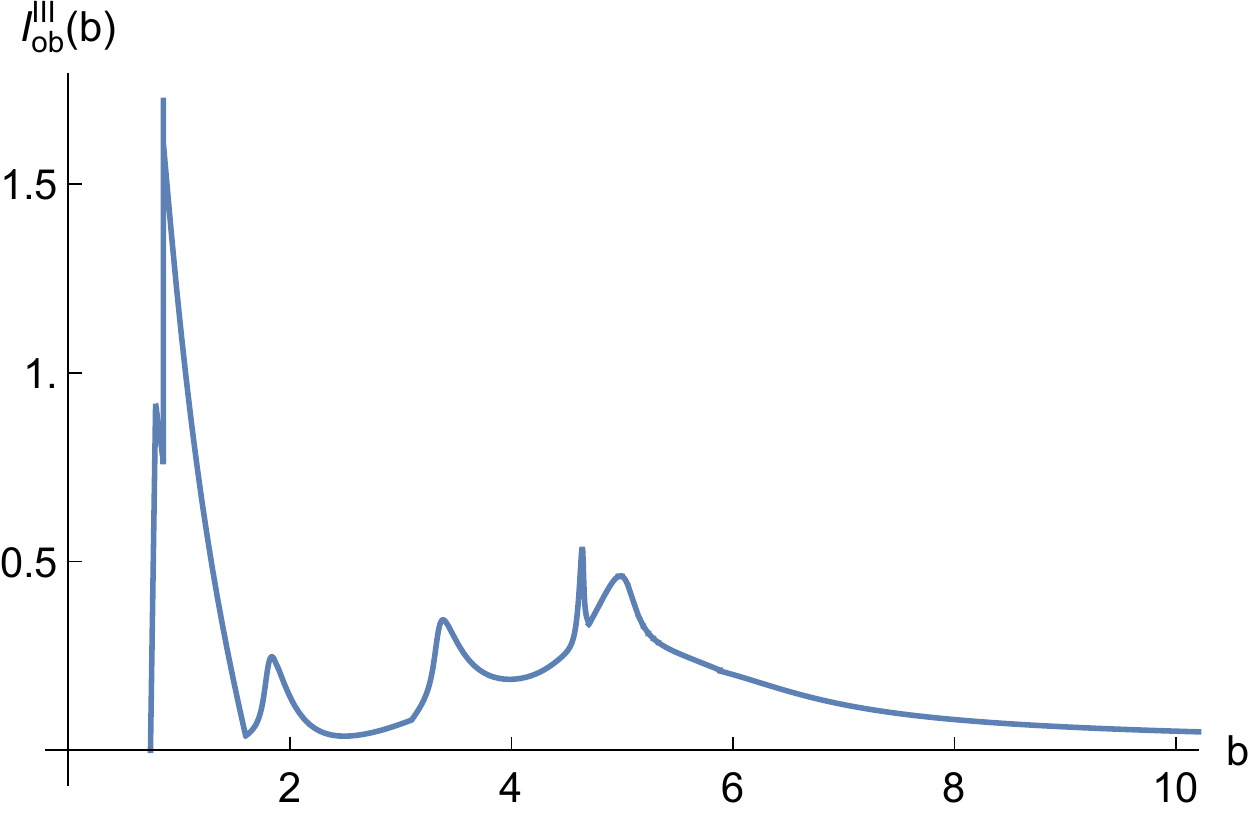}
\includegraphics[width=5.9cm,height=5.2cm]{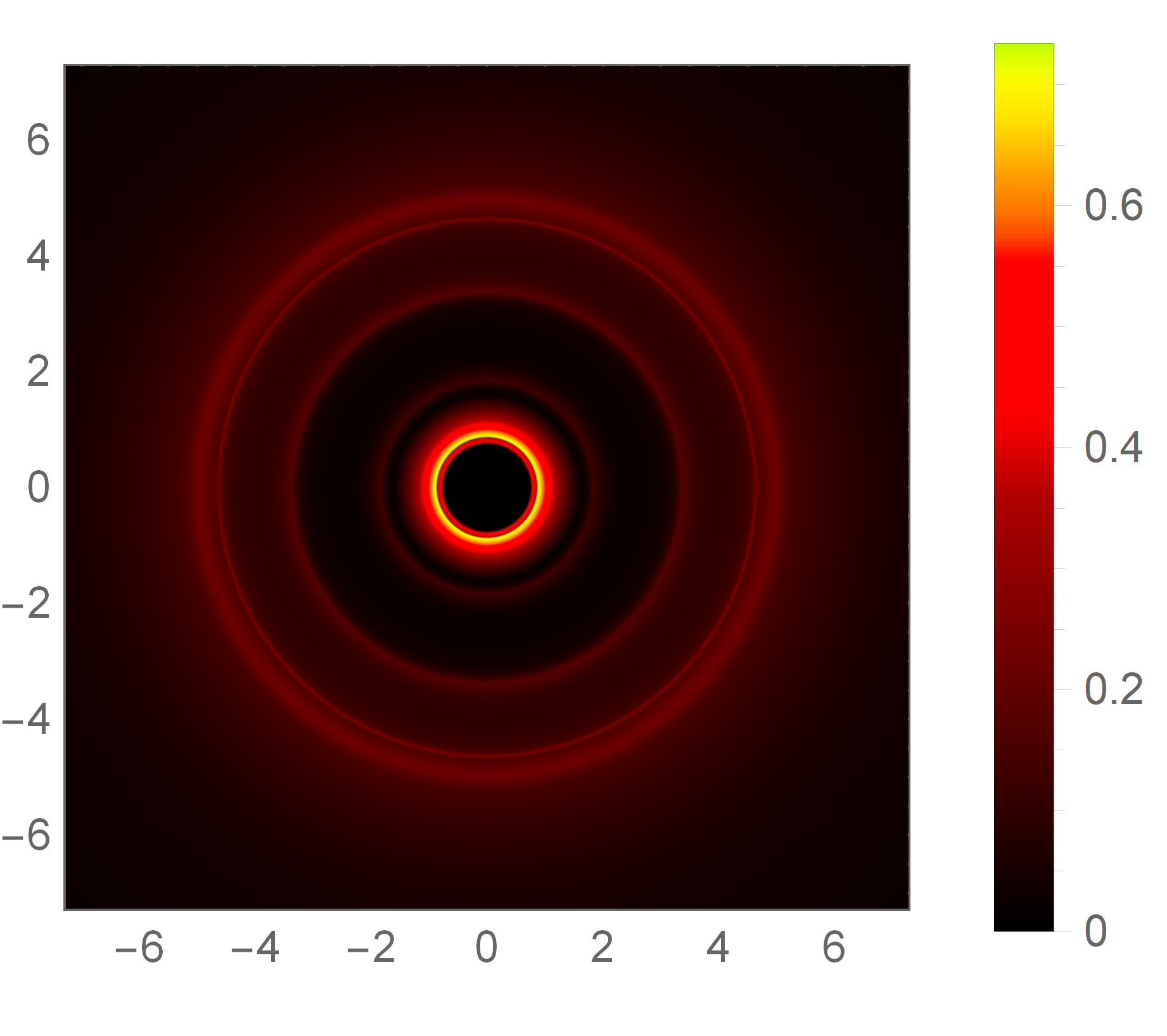}
\caption{The emitted luminosity (left), the observed one (middle) and the optical appearance (right) for the traversable wormhole (in units of $M=1$) having two photon spheres, $a=6/7$, for the Model I (top figures), Model II (middle figures) and Model III (bottom figures).}
\label{fig:shadowWHcase}
\end{figure*}

Let us consider now the case of traversable wormholes having two photon spheres, illustrated here by the choice $a=6/7$. In Fig. \ref{fig:shadowWHcase} we depict the emitted profile (left panels), the observed one (middle), and the optical appearances (right) for this case with the three Models introduced above (top, middle, and bottom, respectively). The most salient features of all these plots are driven by the fact that the presence of the second photon sphere at $b_c^1$ induces three new sources of (observed) emission in the region $b_c^1 \lesssim b<b_c^2$, which yields much clearer and succulent optical appearances.

These new features are clearly seen in Model I (top figures), where again the fact that the inner edge of the disk in the emitted luminosity only extends up to the innermost stable circular orbit for time-like observers allows one to see the isolated presence of three additional spikes in the inner part of the observed luminosity. These are associated to the (starting from the innermost one) photon ring and lensing emission of the inner photon sphere, and to the photon ring emission of the inner part of the outer photon sphere.  These three new spikes produce the presence of three additional (concentric) light rings in the inner region of the associated optical appearance of the object (besides the usual three already found for the Model I when considering the two-horizons configurations previously described). This fact is in agreement with what we expected from the ray-tracing analysis performed in Sec. \ref{sec:Ray}. Moreover, as already pointed out by the almost vertical slope of the corresponding transfer function in the right panel of Fig. \ref{fig:transfunc}, the luminosity of the innermost additional light ring is so dimmed (roughly one part in $10^6$), which is far beyond what can be seen at naked eye. However, the other two additional light rings are clearly visible, storing $\sim 2.5\%$ and $\sim 2.8\%$ of the total luminosity budget. Indeed, these two contributions are larger than the one associated to the standard photon ring of the outer critical curve, which amounts to $\sim 0.5\%$. The presence of these additional light rings and their associated luminosities are trademarks of this kind of compact object having two critical curves, and could potentially be able to act as smoking guns of the existence of any of such objects in the cosmic zoo.

In Model II (middle figures), we again have six peaks of observed intensity, including two additional photon ring emissions and a lensing one driven by the presence of the inner critical curve, though now we have some overlapping of different emissions. From the innermost to the outermost the spike of the first of the new photon ring emissions and the new lensing contribution are clearly seen isolated, followed by the dominance and smooth fall off of the direct emission. Next another spike by the new photon ring emissions is superimposed with the direct one before quickly going off, and finally the lensed and photon ring emissions associated to the outer critical curve appear superimposed (at almost the same impact parameter) with the direct emission. In the corresponding optical appearance plot, the innermost photon ring is again suppressed by a factor $\sim 10^{-6}$ and therefore not visible, while next to it a lensing ring accounting for a $\sim 9.2\%$ share of the total luminosity  is apparent. Going to larger impact parameters, the direct emission (plus the additional photon ring contribution) yield a wide ring of radiation, and in the outermost part another thin ring made up of the lensing and photon ring emissions of the outer critical curve appears, accounting for a $\sim 13.6\%$ of the total luminosity budget. Note that in this Model II the contribution of the direct emission to the total luminosity is significantly decreased up to a fraction of roughly $2/3$ of the total luminosity (compare this with the $\sim 90\%$ in the Schwarzschild case).

In Model III (bottom figures) things get very messy given the fact that the combination of direct, lensed, and photon ring emissions driven by the two critical curves has a complicated pattern of peaks and intermediate valleys in the observed intensity, which never goes completely to zero. Despite this, one can still identify six different peaks associated to the six contributions. The  two innermost peaks are a combination of the direct/lensed/photon ring emissions attaining the maximum of the luminosity, which revels itself in the optical appearance plot as a bright ring of radiation extending all the way down to the inner blank shadow of the wormhole, $b_{is}$. As we move away to larger impact factors, the chain of peaks manifests as additional blurred rings in the optical appearance, with intermediate dark-brown regions rather than totally black associated to the non-vanishing valleys of observed intensity. Moreover, since the peaks have a relatively low height except at the innermost region, this manifests as a large contrast between this region (containing up to a $\sim 85\%$ of the total luminosity) and the rest of the optical appearance of the object.

The bottom line of the discussion above is that the presence of a second (accessible) critical curve in the geometrical background of the traversable wormholes generically yields three additional contributions (as long as we consider direct/lensed/photon ring trajectories) to the observed intensity, whose role in the distribution of the total luminosity is heavily influenced by the emission features of the accretion disk model chosen. Indeed, the corresponding optical appearance is terribly different depending on such a model, since the overlapping between the different contributions to the total luminosity may completely change the distribution and width of the light rings associated to the different trajectories. In any case, for all three models the optical appearance bears little resemblance to the black hole ones studied in the previous section, therefore supporting the conclusion on the feasibility of these two critical curves to act as clear observational discriminators between this type of traversable wormholes and standard black hole solutions.

\section{Conclusion} \label{sec:V}

Testing the Kerr hypothesis against its many competitors has been a relevant trend in the last few years, particularly after the beginning of multimessenger astronomy. In this sense, the year 2019 brought to our weaponry the analysis of black hole shadows, namely, the central depression to a bright ring of radiation bounded by a critical curve (an unstable photon sphere), a naive picture that must be significantly upgraded when the features of the main source of illumination is provided by the accretion disk surrounding the black hole (or its mimicker).

In this work we have studied the shadows, light rings, and optical appearances of a family of spherically symmetric solutions that asymptotes to the Schwarzschild space-time, when surrounded by an optically thin (i.e. transparent to its own radiation) accretion disk. Such a family interpolates between the Schwarzschild solution, a family of two-horizons black holes, an extreme black hole, and a family of traversable wormhole solutions depending on a single parameter. The corresponding solutions have a number of interesting properties, including a subfamily of the traversable wormhole configurations having two critical curves. We constructed the ray-tracing for the two-horizons black holes and the latter configurations, which corresponds to the backtrack of the geodesic equation from the observer's screen to the point of the sky (or to the inner region of the object) a given light ray came from. This allowed us to identify the range of impact parameters contributing to the direct (light rays deflected less than 90 degrees), lensed (two intersections with the equatorial plane) and photon ring  (three intersections) emissions, assuming as negligible other emissions intersecting the equatorial plane more times.

Once the impact parameter region was neatly identified, we considered a geometrically thin accretion disk as the main source of illumination, and used three analytical canonical toy models of the disk whose inner edge (and effective source of emission) extends up to three relevant surfaces: the innermost stable circular orbit for time-like observers, the (outer, if more than one) critical curve itself, and the event horizon (in the black hole case) or the wormhole throat (in the wormhole case). The observed emission corresponds to a gravitational redshift of the emitted one, bearing in mind the different intersections with the disk in the direct/lensed/photon ring trajectories and the additional intensities they pick up depending on the emission profile of the disk. We thus found the corresponding optical appearances, which show significant differences depending on such an emission profile. In the two-horizons black hole case the main difference as compared with the Schwarzschild solution is a moderate enhancement of the width and luminosity of the light rings. However, for the traversable wormhole with the presence of two photon spheres, the existence of new lensed/photon ring contributions associated to the inner critical curve yields three additional peaks in the observed luminosity in such a way that the modelling of the disk has a much more dramatic impact on the corresponding optical appearances. Indeed, the three images of Fig. \ref{fig:shadowWHcase} barely resemble each other. This fact further supports how strongly does the optical appearance of these objects depend on the details of the accretion disk, and how once an accretion model is chosen, the corresponding images deviate dramatically from the black hole (Schwarzschild) counterparts.

In view of the changes in the sizes and locations as well as on the contributions to the total luminosity of the different light rings induced by the traversable wormhole configurations, they are hardly a viable candidate to represent the supermassive object found at the center of the M87 galaxy by the EHT Collaboration, despite the low-resolution of the images available so far. The two-horizon black holes, on the contrary, yield mild modifications to the Schwarzschild predictions, which however would difficult their detectability.  In any case, the simple model presented here illustrates the existence of qualitatively new features in some black hole mimickers having two critical curves, which can be used either as smoking guns of the existence of new Physics, or to rule them out as viable alternatives to the Schwarzschild (Kerr) solution. Investigating this kind of observational discriminators among compact objects that asymptote to the same space-time is of timely interest, should the Kerr solution happen not to describe every single (ultra)-compact object in the universe. In particular, very long baseline interferometry could be able to search for the existence of additional light rings by resolving out diffuse but sharp fluxes in an image \cite{Johnson:2019ljv,Aratore:2021usi}, and therefore the detection of such features - should they be present - could be within reach of the next generation of interferometers.

The results found in this paper point at the need to carry on with the search for clear and clean observational discriminators for the existence of non-GR black holes or black hole mimickers, such as the double critical curves investigated here. This can be done either on a case-by-case basis by finding new such compact objects from a well defined theory of gravity (i.e. either within GR or beyond of it) and matter action, or via parametric deviations of the Kerr black hole shadow in a theory-agnostic way, see e.g. \cite{Medeiros:2019cde,Nampalliwar:2021oqr,Bronnikov:2021liv}. A systematic implementation of this issue would require to investigate every consistent and well-defined shape of a spherically symmetric metric, incorporate rotation, consider moderate inclinations of the disk with respect to the observer's screen, and refine our modelling of the accretion's disk optical and geometrical properties as well as their dynamical behaviour. Therefore, the challenge to confirm further or to refute the Kerr hypothesis is more alive than ever and we hope to further report our progress on this issue soon.

\section*{Acknowledgements}

MG is funded by the predoctoral contract 2018-T1/TIC-10431 and acknowledges further support by the European Regional Development Fund under the Dora Plus scholarship grants.  DRG is funded by the {\it Atracci\'on de Talento Investigador} programme of the Comunidad de Madrid (Spain) No. 2018-T1/TIC-10431. DS-CG is funded by the University of Valladolid (Spain), Ref. POSTDOC UVA20.  This work is supported by the Spanish Grants FIS2017-84440-C2-1-P, PID2019-108485GB-I00, PID2020-116567GB-C21 and PID2020-116567GB-C21 funded by MCIN/AEI/10.13039/501100011033 (``ERDF A way of making Europe" and ``PGC Generaci\'on de Conocimiento"), the project PROMETEO/2020/079 (Generalitat Valenciana), the project H2020-MSCA-RISE-2017 Grant FunFiCO- 777740, the project i-COOPB20462 (CSIC),  the FCT projects No. PTDC/FIS-PAR/31938/2017 and PTDC/FIS-OUT/29048/2017, and the Edital 006/2018 PRONEX (FAPESQ-PB/CNPQ, Brazil, Grant 0015/2019). This article is based upon work from COST Action CA18108, supported by COST (European Cooperation in Science and Technology). All images of this paper were obtained with our own codes implemented within Mathematica\circledR.


\begin{thebibliography}{00}

\bibitem{Robinson:1975bv}
D.~C.~Robinson,
Phys. Rev. Lett. \textbf{34} (1975) 905.

\bibitem{Zajacek:2019kla}
M.~Zaja\v{c}ek and A.~Tursunov,
[arXiv:1904.04654 [astro-ph.GA]].

\bibitem{Hawking:1971ei}
S.~Hawking,
Mon. Not. Roy. Astron. Soc. \textbf{152} (1971) 75.

\bibitem{Senovilla:2014gza}
J.~M.~M.~Senovilla and D.~Garfinkle,
Class. Quant. Grav. \textbf{32} (2015)  124008.

\bibitem{Curiel}
E. Curiel, ``Singularities and black holes", in the \href{ https://plato.stanford.edu/entries/spacetime-singularities/}{Stanford Encyclopedia of Philosophy}.

\bibitem{Psaltis:2020ctj}
D.~Psaltis, C.~Talbot, E.~Payne and I.~Mandel,
Phys. Rev. D \textbf{103} (2021) 104036.

\bibitem{Abramowicz:2002vt}
M.~A.~Abramowicz, W.~Kluzniak and J.~P.~Lasota,
Astron. Astrophys. \textbf{396} (2002) L31.

\bibitem{Glampedakis:2021oie}
K.~Glampedakis and G.~Pappas,
Phys. Rev. D \textbf{104} (2021) L081503.

\bibitem{Shashank:2021giy}
S.~Shashank and C.~Bambi,
[arXiv:2112.05388 [gr-qc]].

\bibitem{Johannsen:2011dh}
T.~Johannsen and D.~Psaltis,
Phys. Rev. D \textbf{83} (2011) 124015.

\bibitem{Konoplya:2016jvv}
R.~Konoplya, L.~Rezzolla and A.~Zhidenko,
Phys. Rev. D \textbf{93} (2016) 064015.



\bibitem{Johnson-Mcdaniel:2018cdu}
N.~K.~Johnson-Mcdaniel, \textit{et al.},
Phys. Rev. D \textbf{102} (2020) 123010.

\bibitem{Simpson:2021dyo}
A.~Simpson and M.~Visser,
[arXiv:2111.12329 [gr-qc]].

\bibitem{Liebling:2012fv}
S.~L.~Liebling and C.~Palenzuela,
Living Rev. Rel. \textbf{15} (2012) 6.

\bibitem{Mazur:2001fv}
P.~O.~Mazur and E.~Mottola,
[arXiv:gr-qc/0109035].

\bibitem{Mathur:2005zp}
S.~D.~Mathur,
Fortsch. Phys. \textbf{53} (2005) 793.

\bibitem{VisserBook}
M. Visser, ``Lorentzian wormholes" (Springer-Verlag, 1995).

\bibitem{Cardoso:2019rvt}
V.~Cardoso and P.~Pani,
Living Rev. Rel. \textbf{22} (2019)  4.

\bibitem{Addazi:2021xuf}
A.~Addazi, \textit{et al.}
[arXiv:2111.05659 [hep-ph]].

\bibitem{SchneiderBook}
P. Schneider, {\it et al.}, ``Gravitational Lensing: Strong, Weak and Micro", Saas-Fee Advanced Course 33 (Springer-Verlag, Berlin-Heidelberg, 2006).

\bibitem{Cunha:2018acu}
P.~V.~P.~Cunha and C.~A.~R.~Herdeiro,
Gen. Rel. Grav. \textbf{50} (2018)  42.


\bibitem{Bozza:2002zj}
V.~Bozza,
Phys. Rev. D \textbf{66} (2002) 103001.

\bibitem{Luminet:1979nyg}
J.~P.~Luminet,
Astron. Astrophys. \textbf{75} (1979) 228.

\bibitem{Falcke:1999pj}
H.~Falcke, F.~Melia and E.~Agol,
Astrophys. J. Lett. \textbf{528} (2000) L13.

\bibitem{Narayan:2019imo}
R.~Narayan, M.~D.~Johnson and C.~F.~Gammie,
Astrophys. J. Lett. \textbf{885} (2019) L33.

\bibitem{Chael:2021rjo}
A.~Chael, M.~D.~Johnson and A.~Lupsasca,
Astrophys. J. \textbf{918} (2021) 6.

\bibitem{Gralla:2019xty}
S.~E.~Gralla, D.~E.~Holz and R.~M.~Wald,
Phys. Rev. D \textbf{100} (2019) 024018.


\bibitem{Akiyama:2019cqa}
K.~Akiyama \textit{et al.} [Event Horizon Telescope],
Astrophys. J. Lett. \textbf{875} (2019) L1.


\bibitem{Wei:2021lku}
S.~W.~Wei and Y.~C.~Zou,
[arXiv:2108.02415 [gr-qc]].

\bibitem{Wielgus:2021peu}
M.~Wielgus,
Phys. Rev. D \textbf{104} (2021) 124058.

\bibitem{Bambi:2017iyh}
C.~Bambi,
Annalen Phys. \textbf{530} (2018) 1700430.

\bibitem{Psaltis:2020lvx}
D.~Psaltis \textit{et al.} [Event Horizon Telescope],
Phys. Rev. Lett. \textbf{125} (2020) 141104.


\bibitem{Held:2019xde}
A.~Held, R.~Gold and A.~Eichhorn,
JCAP \textbf{06} (2019) 029.

\bibitem{Shaikh:2019hbm}
R.~Shaikh and P.~S.~Joshi,
JCAP \textbf{10} (2019) 064.

\bibitem{Zeng:2020vsj}
X.~X.~Zeng and H.~Q.~Zhang,
Eur. Phys. J. C \textbf{80} (2020) 1058.

\bibitem{Paul:2019trt}
S.~Paul, R.~Shaikh, P.~Banerjee and T.~Sarkar,
JCAP \textbf{03} (2020) 055.

\bibitem{Qin:2020xzu}
X.~Qin, S.~Chen and J.~Jing,
Class. Quant. Grav. \textbf{38} (2021)  115008.

\bibitem{Li:2021riw}
G.~P.~Li and K.~J.~He,
JCAP \textbf{06} (2021) 037.

\bibitem{Shaikh:2021cvl}
R.~Shaikh, S.~Paul, P.~Banerjee and T.~Sarkar,
[arXiv:2105.12057 [gr-qc]].

\bibitem{Zeng:2020dco}
X.~X.~Zeng, H.~Q.~Zhang and H.~Zhang,
Eur. Phys. J. C \textbf{80} (2020) 872.

\bibitem{He:2021htq}
K.~J.~He, S.~Guo, S.~C.~Tan and G.~P.~Li,
[arXiv:2103.13664 [hep-th]].

\bibitem{Gan:2021pwu}
Q.~Gan, P.~Wang, H.~Wu and H.~Yang,
Phys. Rev. D \textbf{104} (2021) 024003.

\bibitem{Eichhorn:2021iwq}
A.~Eichhorn and A.~Held,
JCAP \textbf{05} (2021) 073.

\bibitem{Guerrero:2021ues}
M.~Guerrero, G.~J.~Olmo, D.~Rubiera-Garcia and D.~S.~C.~G\'omez,
JCAP \textbf{08} (2021) 036.

\bibitem{Liu:2021yev}
C.~Liu, S.~Yang, Q.~Wu and T.~Zhu,
[arXiv:2107.04811 [gr-qc]].

\bibitem{Okyay:2021nnh}
M.~Okyay and A.~\"Ovg\"un,
JCAP \textbf{01} (2022) 009.

\bibitem{Afrin:2021wlj}
M.~Afrin and S.~G.~Ghosh,
[arXiv:2110.05258 [gr-qc]].

\bibitem{Zeng:2021mok}
X.~X.~Zeng, K.~J.~He and G.~P.~Li,
[arXiv:2111.05090 [gr-qc]].

\bibitem{Guo:2021wid}
Y.~Guo and Y.~G.~Miao,
[arXiv:2112.01747 [gr-qc]].

\bibitem{Gan:2021xdl}
Q.~Gan, P.~Wang, H.~Wu and H.~Yang,
Phys. Rev. D \textbf{104} (2021) 044049.

\bibitem{Perlick:2021aok}
V.~Perlick and O.~Y.~Tsupko,
Phys. Rept. \textbf{947} (2022) 1.

\bibitem{Lara:2021zth}
G.~Lara, S.~H.~V\"olkel and E.~Barausse,
Phys. Rev. D \textbf{104} (2021)  124041.

\bibitem{Psaltis}
D. Psaltis, Gen. Rel. Grav. \textbf{51} (2019) 137.


\bibitem{Olmo:2021piq}
G.~J.~Olmo, D.~Rubiera-Garcia and D.~S.~C.~G\'omez,
[arXiv:2110.10002 [gr-qc]].


\bibitem{Wielgus:2020uqz}
M.~Wielgus, J.~Horak, F.~Vincent and M.~Abramowicz,
Phys. Rev. D \textbf{102} (2020) 084044.

\bibitem{Wang:2020emr}
X.~Wang, P.~C.~Li, C.~Y.~Zhang and M.~Guo,
Phys. Lett. B \textbf{811} (2020) 135930.

\bibitem{Guerrero:2021pxt}
M.~Guerrero, G.~J.~Olmo and D.~Rubiera-Garcia,
JCAP \textbf{04} (2021) 066.

\bibitem{Peng:2021osd}
J.~Peng, M.~Guo and X.~H.~Feng,
Phys. Rev. D \textbf{104} (2021) 124010.

\bibitem{Shaikh:2019jfr}
R.~Shaikh, P.~Banerjee, S.~Paul and T.~Sarkar,
JCAP \textbf{07} (2019) 028.

\bibitem{Shaikh:2019itn}
R.~Shaikh, P.~Banerjee, S.~Paul and T.~Sarkar,
Phys. Rev. D \textbf{99} (2019) 104040.

\bibitem{Lobo:2020ffi}
F.~S.~N.~Lobo, M.~E.~Rodrigues, M.~V.~d.~S.~Silva, A.~Simpson and M.~Visser,
Phys. Rev. D \textbf{103} (2021) 084052.

\bibitem{Tsukamoto:2021caq}
N.~Tsukamoto,
Phys. Rev. D \textbf{104} (2021) 064022.


\bibitem{Ansoldi:2008jw}
S.~Ansoldi,
[arXiv:0802.0330 [gr-qc]].

\bibitem{Maeda:2021jdc}
H.~Maeda,
[arXiv:2107.04791 [gr-qc]].

\bibitem{Bardeen}
J, Bardeen, presented at GR5, Tiflis, U.S.S.R., and published in the conference proceedings in the U.S.S.R. (1968).

\bibitem{Bronnikov:2021uta}
K.~A.~Bronnikov and R.~K.~Walia,
[arXiv:2112.13198 [gr-qc]].

\bibitem{Cardoso:2014sna}
V.~Cardoso, L.~C.~B.~Crispino, C.~F.~B.~Macedo, H.~Okawa and P.~Pani,
Phys. Rev. D \textbf{90} (2014) 044069.

\bibitem{Bisnovatyi-Kogan:2022ujt}
G.~S.~Bisnovatyi-Kogan and O.~Y.~Tsupko,
[arXiv:2201.01716 [gr-qc]].

\bibitem{Gralla:2019drh}
S.~E.~Gralla and A.~Lupsasca,
Phys. Rev. D \textbf{101} (2020) 044031.

\bibitem{Vincent:2020dij}
F.~H.~Vincent, \textit{et al.},
Astron. Astrophys. \textbf{646} (2021) A37.

\bibitem{RadiativeBook}
George B. Rybicki, Alan P. Lightman, ``Radiative Processes in Astrophysics" (New York: Wiley-VCH, 2004).


\bibitem{Gold:2020iql}
R.~Gold, \textit{et al.}
Astrophys. J. \textbf{897} (2020) 148.

\bibitem{Johnson:2019ljv}
M.~D.~Johnson, \textit{et al.}
Sci. Adv. \textbf{6} (2020) eaaz1310.

\bibitem{Aratore:2021usi}
F.~Aratore and V.~Bozza,
JCAP \textbf{10} (2021) 054.


\bibitem{Medeiros:2019cde}
L.~Medeiros, D.~Psaltis and F.~\"Ozel,
Astrophys. J. \textbf{896} (2020)  7.

\bibitem{Nampalliwar:2021oqr}
S.~Nampalliwar and S.~K,
[arXiv:2108.01190 [gr-qc]].

\bibitem{Bronnikov:2021liv}
K.~A.~Bronnikov, R.~A.~Konoplya and T.~D.~Pappas,
Phys. Rev. D \textbf{103} (2021) 124062.

\end{thebibliography}
\end{document}